\documentclass[useAMS,usenatbib]{mn2e}
\usepackage[english]{babel}
\usepackage{fancyhdr}

\usepackage{epsfig}
\usepackage{subfigure}

\usepackage{times}
\usepackage{natbib}

\newif\ifAMStwofonts
\AMStwofontstrue

\newcommand{\be}{\begin{equation}}
\newcommand{\ee}{\end{equation}}
\newcommand{\ba}{\begin{eqnarray}}
\newcommand{\ea}{\end{eqnarray}}
\newcommand{\brr}{\begin{array}}
\newcommand{\err}{\end{array}}
\newcommand{\bc}{\begin{center}}
\newcommand{\ec}{\end{center}}

\newcommand{\mincir}{\raise
  -2.truept\hbox{\rlap{\hbox{$\sim$}}\raise5.truept \hbox{$<$}\ }}
\newcommand{\magcir}{\raise
  -2.truept\hbox{\rlap{\hbox{$\sim$}}\raise5.truept \hbox{$>$}\ }}
\newcommand{\siml}{\raise
  -2.truept\hbox{\rlap{\hbox{$\sim$}}\raise5.truept \hbox{$<$}\ }}
\newcommand{\simg}{\raise
  -2.truept\hbox{\rlap{\hbox{$\sim$}}\raise5.truept \hbox{$>$}\ }}

\title[How does gas cool in DM halos?] {How does gas cool in DM
  halos?}  \author[Viola et al.] {M. Viola$^{1}$,
  P. Monaco$^{1,2}$, S. Borgani$^{1,2,3}$, G. Murante$^{4}$ \&
  L. Tornatore$^{1,5}$ \\~\\
  $^1$ Dipartimento di Astronomia dell'Universit\`a di Trieste, via
  Tiepolo 11, I-34131 Trieste, Italy
  (viola,borgani,monaco,tornatore@oats.inaf.it)\\ $^2$ INAF -- Istituto
  Nazionale di Astrofisica, Trieste, Italy\\ $^3$ INFN -- Istituto
  Nazionale di Fisica Nucleare, Trieste, Italy\\ $^4$ INAF --
  Osservatorio Astronomico di Torino, Strada Osservatorio 20, I-10025
  Pino Torinese, Italy (murante@oato.inaf.it)\\ 
  $^5$ SISSA -- via Beirut 4, I-34014 Trieste, Italy \\ }

\begin{document}
\label{firstpage}
\maketitle

\begin{abstract}
In order to study the process of cooling in dark-matter (DM) halos and
assess how well simple models can represent it, we run a set of
radiative SPH hydrodynamical simulations of isolated halos, with gas
sitting initially in hydrostatic equilibrium within
Navarro-Frenk-White (NFW) potential wells. Simulations include
radiative cooling and a scheme to convert high density cold gas
particles into collisionless stars, neglecting any astrophysical
source of energy feedback. After having assessed the numerical
stability of the simulations, we compare the resulting evolution of
the cooled mass with the predictions of the classical cooling model of
White \& Frenk and of the cooling model proposed in the {\sc morgana}
code of galaxy formation. We find that the classical model predicts
fractions of cooled mass which, after about two central cooling times,
are about one order of magnitude smaller than those found in
simulations. Although this difference decreases with time, after 8
central cooling times, when simulations are stopped, the
difference still amounts to a factor of 2--3.  We ascribe this
difference to the lack of validity of the assumption that a mass shell
takes one cooling time, as computed on the initial conditions, to cool
to very low temperature. Indeed, we find from simulations that cooling
SPH particles take most time in traveling, at roughly constant
temperature and increasing density, from their initial position to a
central cooling region, where they quickly cool down to $\sim10^4$ K.
We show that in this case the total cooling time is shorter than that
computed on the initial conditions, as a consequence of the stronger
radiative losses associated to the higher density experienced by these
particles. As a consequence the mass cooling flow is stronger than
that predicted by the classical model.

The {\sc morgana} model, which computes the cooling rate as an
integral over the contribution of cooling shells and does not make
assumptions on the time needed by shells to reach very low
temperature, better agrees with the cooled mass fraction found in the
simulations, especially at early times, when the density profile of
the cooling gas is shallow. With the addition of the simple assumption
that the increase of the radius of the cooling region is counteracted
by a shrinking at the sound speed, the {\sc morgana} model is also
able to reproduce for all simulations the evolution of the cooled mass
fraction to within 20--50 per cent, thereby providing a substantial
improvement with respect to the classical model.  Finally, we
provide a very simple fitting function which accurately reproduces the
cooling flow for the first $\sim10$ central cooling times.
\end{abstract}


\section{Introduction}
Understanding galaxy formation is one of the most important challenges
of modern cosmology. The rather stringent constraints on cosmological
parameters now placed by a number of independent observations
\citep[e.g.,][ for a recent review]{Springel06} allows us to precisely
set the initial conditions from which the formation of cosmic
structures has started. As a consequence, understanding the complex
astrophysical processes, related to the evolution of the baryonic
component, represents now the missing link toward a successful
description of galaxy formation and evolution.  So far, two
alternative approaches have been pursued to make quantitative
predictions on the observational properties of the galaxy population
and their evolution in the cosmological context. The first one is
based on the so-called semi-analytical models (SAMs, hereafter; e.g.,
\citealt{Kauffmann93,Somerville99,Cole00,Menci05,Monaco07}). In this
approach, the background cosmological model predicts the
hierarchical build-up of the Dark Matter (DM) halos, where gas flows
in, cools and gives rise to the formation of galaxies, while the
complex interplay between gas cooling, star formation, chemical
enrichment and release of energy from supernova (SN) explosions and
Active Galactic Nuclei (AGN) is modeled through a set of simplified or
phenomenological models, which are specified by a number of free
parameters. A posteriori, the values of the relevant parameters should
then be constrained by comparing SAM predictions to observational
data. The rather low computational cost of this approach makes it a
quite flexible tool to explore the model parameter space.

The second approach is based on resorting to full hydrodynamical
simulations, which include the processes of gas cooling and suitable
sub-resolution recipes for star formation and feedback. The obvious
advantage of this method, with respect to SAMs, is that galaxy
formation can be described by following in detail the gas-dynamical
processes which determine the evolution of the cosmic baryons during
the shaping of the large-scale cosmic structures. However, its
limitation lies in the high computational cost, which makes it
difficult to explore in detail the parameter space describing
the process of galaxy formation and evolution. For this reason,
following galaxy formation with full hydrodynamical simulations in a
cosmological environment of several tens of Mpc is a challenging task
for simulations of the present generation \citep[e.g.,
][]{Nagamine04,Nagai05,Romeo05,Saro06}.

This discussion shows that SAMs and hydrodynamical simulations provide
complementary approaches to the cosmological study of galaxy
formation. The ability of hydrodynamical simulations of accurately
following gas dynamics calls for the need of a close comparison
between these two approaches, in order to test the basic assumptions
of the SAMs. The best regime to perform this comparison is when one
excludes the effect of all those processes, like feedback in energy
and metals, whose different modeling in SAMs and simulation codes
would make the comparison scarcely telling. Since gas cooling is the
most basic ingredient in any model of galaxy formation \citep[e.g.,
][]{White91}, an interesting comparison would be performed when
cooling is the only process turned on. In this spirit \cite{Benson01},
using a hydrodynamical simulation of a cosmological box and a
stripped-down version of SAM, compared the statistical properties of
``galaxies'' found in the two cases. They discovered that SPH
simulation and SAM give similar results for the thermodynamical
evolution of gas and that there is a very good agreement in terms of
final fractions of hot, cold and uncollapsed gas.

Similar conclusions were reached by \cite{Helly03} and
\cite{Cattaneo07}. They improved the comparison performed by
\cite{Benson01} by giving to the down-stripped SAM the same halo
merger histories extracted from the cosmological simulation. In this
way they were able to compare cooling in DM halos not only
statistically but on an object-by-object basis.  The result was again
that the two methods provide comparable ``galaxy'' populations.
\cite{Yoshida02} performed a similar comparison for a simulation of a
single galaxy cluster, obtaining similarly good results.

While the general agreement between the two methods is encouraging,
still all the above analyses generally concentrated on comparing the
statistical properties of the galaxy populations. Furthermore, if one
wants to test the reliability of the cooling model implemented in the
SAMs, the cleanest approach would be that of turning off the
complications associated to the hierarchical merging of halos, thereby
allowing gas to cool in isolated halos.

The purpose of this paper is to present a detailed comparison between
the predictions of cooling models, as implemented in SAMs, and results
of hydrodynamical simulations in which gas is allowed to cool inside
isolated halos. Our controlled numerical experiments will be run for
halos having the density profile (for DM particles) of \cite{NFW} (NFW
hereafter), with a range of masses, concentration parameters and
average densities (related to the halos' redshift). As a baseline
model for gas cooling, we consider the classical one, as originally
proposed by \cite{White91}.  In this model, the cooling radius is
defined as the radius at which the cooling time equals the time
elapsed since radiative cooling is turned on.  As a result, the growth
rate of the cooled gas mass is simply related to the growth rate of
the cooling radius. This model was claimed by \cite{White91} to be
close the exact self-similar solutions of cooling flows presented by
\cite{Bert89}.  Simulation results will also be compared to another
model of gas cooling, which has been recently proposed by
\cite{Monaco07} in the context of the {\sc morgana} SAM, and is based
on a ``dynamical'' definition of cooling radius. As a result of our
analysis, we will show that the gas cooling rate in the simulations is
initially faster than predicted by the classical cooling model. When
the simulations are stopped, after about 8 central cooling times,
this initial transient causes the classical cooling model to
underestimate the cooled mass by an amount which can be as large as a
factor of three, depending on the halo concentration, density and
mass. A much better agreement with simulations is achieved by the
alternative {\sc morgana} model.

The plan of the paper is as follows. In section 2 we describe first
the ``classical'' analytic model for cooling \citep{White91}, and the
alternative {\sc morgana} cooling model \citep{Monaco07}. In section 3
we present numerical simulations, performed with the {\tt GADGET-2}
code and in section 4 we discuss the results obtained by comparing
simulations to analytical cooling models. We discuss our results in
section 5 and draw our final conclusions in section 6.  A more
technical discussion on the differences between the classical and the
{\sc morgana} cooling models is provided in Appendix A, while
Appendix B gives a very simple fitting formula for the cooling flows.

\section{Analytic models for gas cooling}
\label{section:analytic}

\subsection{Hydrostatic equilibrium in an NFW halo}
\label{section:profile}

All our tests start from a spherical dark matter halo with an NFW
density profile,
\begin{equation}\label{nfw}
\rho (r)=\rho_{\rm crit}\frac{\delta_{c}}{(r/r_{s})(1+r/r_{s})^{2}}\,,
\end{equation}
where $r_{s}$ is a scale radius, $\delta_{c}$ is a characteristic
(dimensionless) density, and $\rho_{\rm crit}$ is the critical cosmic
density.  The gas is assumed to be in hydrostatic equilibrium.  The
equilibrium solution for the gas can be found \citep{Suto98} assuming
that the baryonic fraction of the halo is negligible and the gas, with
density $\rho_g$, temperature $T_g$ and pressure $P_g$, follows a
polytropic equation of state with index $\gamma_p$, $P_g \propto
\rho_g^{\gamma_p}$.
The equation of hydrostatic equilibrium is:
\be
\frac{dP_g}{dr} = -G\frac{\rho_g M(<r)}{r^2}\,.
\label{eq:hydro} \ee 
Here $M(<r)$ is the halo mass within $r$ (we will call
$M_{200}=M(<r_{200})$ the mass within the radius $r_{200}$
encompassing an overdensity of $200\rho_{\rm crit}$), which is given by
the NFW profile. Under these assumptions the equation can be solved
analytically \citep{Komatsu01}:
\begin{eqnarray}
\rho_g(r)  &=& \rho_{g0}\left[1-a\left(1-\frac{\ln(1+c_{\rm nfw}x)}{c_{\rm nfw}x}
\right)\right]^{1/(\gamma_p-1)}\,; \nonumber\\
P_g(r)     &=& P_{g0}\left[1-a\left(1-\frac{\ln(1+c_{\rm nfw}x)}{c_{\rm nfw}x}
\right)\right]^{\gamma_p/(\gamma_p-1)}\,; \label{eq:hydro_sol} \\
T_g(r)     &=& T_{g0}\left[1-a\left(1-\frac{\ln(1+c_{\rm nfw}x)}{c_{\rm nfw}x}
\right)\right]\,. \nonumber 
\end{eqnarray}
Here $\rho_{g0}$, $T_{g0}$ and $P_{g0}$ are the gas density,
temperature and pressure at $r=0$ and $x=r/r_{200}$.  Furthermore,
$c_{\rm nfw}=r_{200}/r_{s}$ is the NFW concentration parameter, while
$\gamma_p$ is the effective polytropic index, which determines the shape
of the temperature profile. The constant $a$ is defined as:
\be
a = \frac{\gamma_p-1}{\gamma_p}\frac{3}{\eta_0}
\frac{c_{\rm nfw}(1+c_{\rm nfw})}{(1+c_{\rm nfw})\ln(1+c_{\rm nfw})-c_{\rm nfw}} \, .
\label{eq:adef} 
\ee
The parameter
\begin{equation}\label{eq:eta0}
\eta_{0}=\frac{3r_{200}k_{B}T_0}{G\mu m_{p}M_{200}}
\end{equation}
carries information about the central gas temperature $T_0$ and the
mass $M_{200}$, thereby fixing the normalization of the polytropic
equation of state. In eq. (\ref{eq:eta0}), $k_B$ is the Boltzmann
constant, $\mu$ is the mean molecular weight ($\simeq 0.58$ for a
plasma of primordial composition), and $m_p$ the proton mass.

Therefore, $\rho_{g0}$ is fixed by the constraint on the total gas
mass $M_g$ within the virial radius.  Calling ${\cal I}$ the integral
\begin{equation}
{\cal I}(\alpha) = \int_{0}^{c_{\rm nfw}} \left[ 1-a\left(1-\frac{\ln(1+t)}{t}
\right)\right]^{\alpha}t^2 dt
\label{eq:integral} \end{equation}
(where for simplicity we declare only the dependence on the $\alpha$
exponent), we have
\begin{equation}
\rho_{g0} =\frac{M_g}{4\pi r_s^3} \times \frac{1}{{\cal I} (1/(\gamma_p-1))}\,.
\label{eq:mass}
\end{equation}
In this way, $\eta_0$ is fixed by the constraint on the total gas
thermal energy $E_g$ within the virial radius:
\begin{equation}
E_g = \frac{6\pi k_ B T_{g0}\rho_{g0}r_s^3}{\mu m_p} 
\times {\cal I} (\gamma_p/(\gamma_p-1))\, .
\label{eq:energy}
\end{equation}
The solution of the two conditions must be found numerically by
iteration.

\subsection{The classical cooling model}
\label{section:classical}

Most SAMs describe the cooling of
gas following the model of \cite{White91}.  The system is
assumed to be in hydrostatic equilibrium at the time $t=0$.  For each
mass shell at radius $r$ a cooling time can be defined as:
\begin{equation}
t_{\rm cool}(r) := \left |{d\ln
T\over dt}\right|^{-1}= \frac{3k_BT_g(r) \mu m_p}
{2\rho_g(r) \Lambda(T_g(r))}\,,
\label{eq:coolingtime}\end{equation}
where $\Lambda$ is the cooling function. The left-hand-side of the
above equation follows from the assumption that $dT$ is computed for
an isobaric transformation. In the following, we assume, for both
simulations and analytical models, $\Lambda(T)$ to be that tabulated
by \cite{SD93} for zero metallicity\footnote{The
cooling rate per unit volume should be $n_e n_i \Lambda$; we transform
the cooling function so that the cooling rate is $n^2\Lambda$, where
$n=n_e+n_i=\rho/\mu m_p$.  For sake of simplicity we do not
make this explicit in the equation.}. The cooling radius at the time $t$
for the classical model, $r_{\rm C}(t)$, is the defined through the
relation
\begin{equation}
r_{\rm C}(t): \ \ \  t_{\rm cool}(r_{\rm C}) = t\,.
\label{eq:coolingradius} \end{equation}
In other words, the function $r_{\rm C}(t)$ is the inverse of the
function $t_{\rm cool}(r)$.  It is then assumed that each shell cools
after one cooling time.  The resulting mass deposition rate reads
\begin{equation}
\dot{M}_{\rm cool} = 4 \pi r^2\rho_g(r_{\rm C}) \frac{dr_{\rm C}}{dt}\, .
\label{eq:classicalcooling} \end{equation}
where a dot denotes a time derivative.

We emphasize that the classical cooling model imposes that a shell of
gas cools exactly after one cooling time, computed on the initial
configuration. The mass cooling rate of equation
(\ref{eq:classicalcooling}), predicted by this model, is generally not
far from that of the self-similar solutions of cooling flows found by
\cite{Bert89}. This point will be further discussed in Section 5 and
in the Appendix.

\subsection{The {\sc morgana} cooling models}
\label{section:morgana}
In the classical cooling model, hot gas is assumed to be located
outside the cooling radius $r_{\rm C}$.  This same
assumption is used in the {\sc morgana} cooling model; we call this
cooling radius $r_{\rm M}$ to distinguish it from the classical one.
The equilibrium gas configuration is computed as in
equations~(\ref{eq:hydro_sol}), but taking into account that no hot mass
is present within $r_{\rm M}$; this implies that the integral
in equation~(\ref{eq:integral}) is evaluated
from $r_{\rm M}/r_s$ to $c_{\rm nfw}$.

Given the equilibrium profile, the cooling rate of a shell of gas of
width $\Delta r$ at a radius $r$ is:
\begin{equation} \Delta \dot{M}_{\rm cool}(r) = 
\frac{4\pi r^2 \rho_g(r)\Delta r}{t_{\rm cool}(r)}\, ,
\label{eq:coolshell} \end{equation}
where $t_{\rm cool}$ is given by equation~(\ref{eq:coolingtime}). The
mass deposition rate is then computed by integrating the contributions of all
the mass shells. In performing this integral, we note that the cooling
time depends on density and temperature, 
the density dependence being always stronger since the temperature
profile is much shallower than the density profile.  The integration
in $r$ can then be performed by assuming $T_g(r)\simeq T_g(r_{\rm
  M})$, while taking into full account the radial dependence of the
density. The resulting total mass deposition rate can then be written as
\begin{eqnarray}\label{eq:morganacooling} 
\lefteqn{\dot{M}_{\rm cool} =}\\&& \frac{4 \pi r_s^3\rho_{g0}}{t_{\rm cool,0}} 
\int_{r_{\rm M}/r_s}^{c_{\rm nfw}} \left[ 1-a\left(1-\frac{\ln(1+t)}{t}
\right)\right]^{2/(\gamma_p-1)}t^2 dt\nonumber.
\end{eqnarray}
We apply this mass cooling rate  starting from $t=t_{\rm cool}(0)$
(the cooling time at $r=0$), under the hypothesis that
nothing cools before that time. The rate of thermal energy
loss by cooling is computed in a similar way:
\begin{eqnarray}\label{eq:coolingenergy} 
\lefteqn{\dot{E}_{\rm cool} = \frac{3kT_g(r_{\rm cool})}{2\mu m_p}
\frac{4 \pi r_s^3\rho_{g0}}{t_{\rm cool,0}}  }\\&&
\times \int_{r_{\rm M}/r_s}^{c_{\rm nfw}} \left[ 1-a\left(1-\frac{\ln(1+t)}{t}
\right)\right]^{2/(\gamma_p-1)}t^2 dt\,.\nonumber
\end{eqnarray}
The lower extreme of integration comes from the assumption that
the hot and cold phases are always separated by a sharp transition,
taking place at the cooling radius $r_{\rm M}$.  By equating the
mass cooled in a time interval $dt$ with the mass contained in a shell
$dr$, one obtains the evolution of the cooling radius:
\begin{equation}
\dot{r}_{\rm M} = \frac{\dot{M}_{\rm cool}}
{4\pi \rho_g(r_{\rm M})r_{\rm M}^2}\,.
\label{eq:drcool1} \end{equation}
The evolution of the system is then followed by numerically
integrating the equations~(\ref{eq:morganacooling}),
(\ref{eq:coolingenergy}) and (\ref{eq:drcool1}), starting after a time
$t_{\rm cool}(0)$.  The integration is performed with a simple
Runge-Kutta integrator with adaptive time-step and the gas profile is
re-computed at each time step.  Clearly, as long as the mass and
thermal energy of each cooled shell is removed from the profile, the
rest of the gas is unperturbed so its profile does not change with
time.

The two main assumptions at the base of the {\sc morgana} cooling
model are: {\bf (i)} all mass shells contribute to cooling according
to equation (\ref{eq:coolshell}), {\bf (ii)} there is anyway a sharp
transition from the hot to the cold phase, which guarantees the
presence of a sharp border separating the regions where cooled and hot
gas resides.

Equation~(\ref{eq:drcool1}) is valid under the assumption that the gas
is pressure-supported at the cooling radius, which is clearly false in
general.  In particular, as long as the time derivative of the cooling
radius is much larger than the gas sound speed, the boundary of the
cooled region propagates so quickly that any gas motion can be
neglected. This holds only at early times and, therefore, for very low
$r_{\rm M}$ values. On the contrary, the late evolution of cooling is better
represented by letting the ``cooling hole''  close at the sound speed.
We then define a further cooling radius $r_{\rm M,ch}$, whose
evolution is given by the equation:
\begin{equation}
\dot{r}_{\rm M,ch} = \frac{\dot{M}_{\rm cool}}
{4\pi \rho_g(r_{\rm M,ch})r_{\rm M,ch}^2} - c_s(r_{\rm M,ch})\,.
\label{eq:drcool2} \end{equation}
Here $c_s$ is the sound speed evaluated at the cooling radius.  This
simple change strongly influences the predictions of the cooling
model.  In fact, the evolution of the cooling radius turns out to be
remarkably different, with $r_{\rm M,ch}<r_{\rm M}$; 
as a consequence, because the gas profile is
recomputed at each time-step, the mass re-distributes over the
available volume, at variance with the previous case in which the gas
profile remains unperturbed beyond $r_{\rm M}$.  We will refer to the
models of equations (\ref{eq:drcool1}) and (\ref{eq:drcool2}) as the
``unclosed'' and ``closed'' {\sc morgana} cooling models,
respectively.

Equation (\ref{eq:drcool2}) clearly gives an oversimplified
description of the process in play, with the implicit assumption that
the gas has time to settle into a new hydrostatic equilibrium
configuration.  However, due to the negative gas temperature profile,
the uncooled gas is predicted to become progressively colder, with the
result that the density and temperature profiles become steeper to
keep satisfying the condition of hydrostatic equilibrium. Eventually,
this leads to a catastrophic cooling involving all the the gas of the
halo.  A simple and effective way to obtain an acceptable behaviour
for $M_{\rm cool}$ is to suppress the sound speed term of equation
(\ref{eq:drcool2}) when the specific thermal energy of the hot gas
becomes smaller than the specific virial energy ($-0.5\, U_H/M_H$,
where $U_H$ is the binding energy of the DM halo). Since this is
admittedly an ad-hoc solution to the above problem, our attitude is
to consider the ``closed'' {\sc morgana} model as an effective model
to describe the evolution of the cooled mass in DM halos.

In summary, we have identified three analytic cooling models:
classical, unclosed {\sc morgana} and closed {\sc morgana}.  The main
difference between the classical and unclosed {\sc morgana} model is
the following: while the classical model equates the time required for
a shell to cool to low temperature with the cooling time computed on
the initial conditions (eq. \ref{eq:coolingtime}), the unclosed {\sc
morgana} model does not rely on this strong assumption.  The main
difference between the unclosed and closed {\sc morgana} models is
that the latter attempts a more realistic description of the evolution
of the cooling radius, taking into account that the cooled gas cannot
provide pressure support to the inflowing cooling gas.

\begin{figure*}
\centerline{
    \epsfig{figure=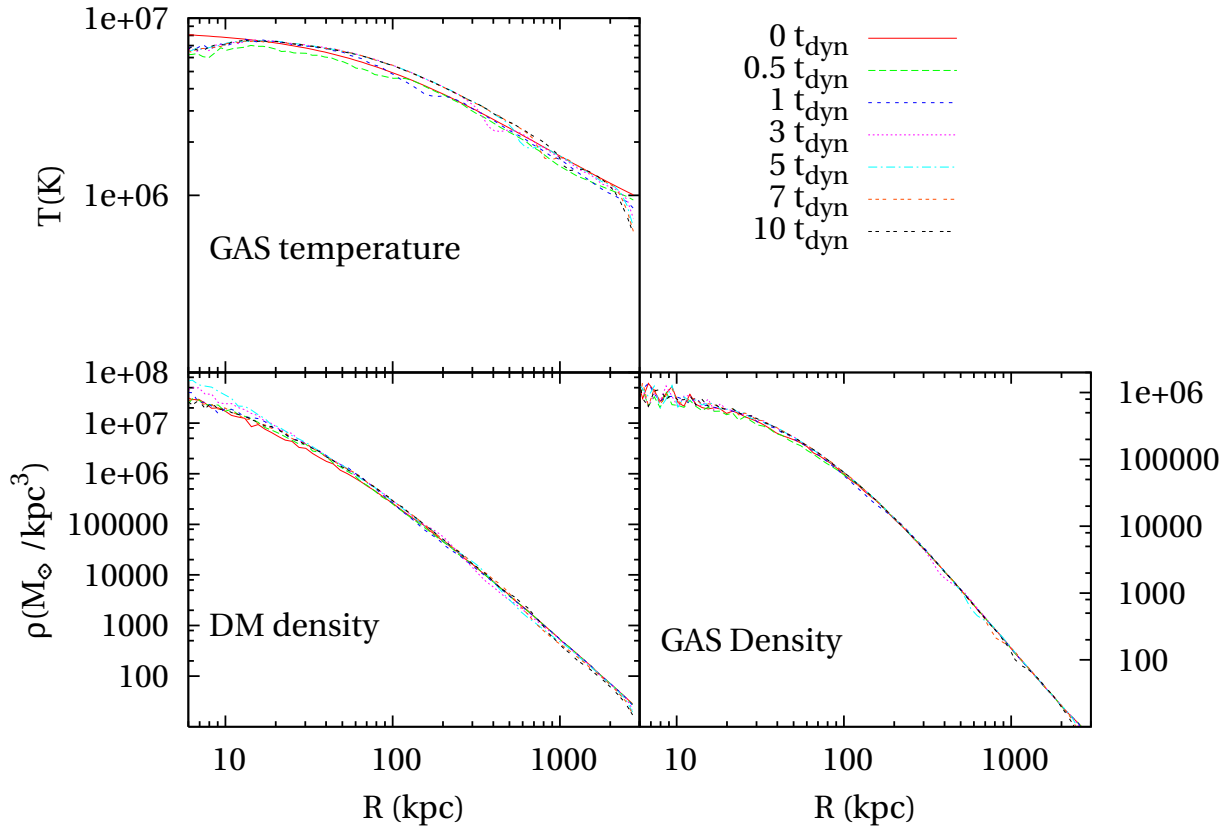,width=12cm,angle=270}}
  \caption{Profiles of gas temperature (\emph{Top panel}), DM density
    (\emph{Bottom left panels}) and gas density (\emph{Bottom right
      panel}) for the non-radiative run of the halo H1, evaluated at 7
    different epochs in the interval
    $[0,10]t_{dyn}$.}\label{fig:relaxed_profile}
\end{figure*}

\section{The simulations}
The simulations that we will discuss in this paper have been obtained
by evolving initial conditions for gas sitting in hydrostatic
equilibrium within isolated halos whose DM density profile is the NFW one. 
They have been performed using {\tt GADGET-2} code, a massively
parallel Tree+SPH code \citep{Springel05gad} with fully adaptive time-step
integration. The version of the code that we used adopts an SPH
formulation with entropy conserving integration and arithmetic
symmetrization of the hydrodynamical forces \citep{SprHern02}, and
includes radiative cooling computed for a primordial plasma with
vanishing metallicity. The above SPH formulation is also that used by
\cite{Yoshida02} in their comparison between SAMs and hydrodynamical
simulations. It ensures the suppression of spurious cooling at the
interfaces between cooled and hot gas \citep[see
also][]{Tornatore03}. In order to follow in detail the trajectories of
gas particles in the phase diagram, while they are undergoing cooling,
we have implemented a quite conservative criterion of time-stepping,
in which the maximum time-step allowed for a gas particle is one
tenth of its cooling time.

In the following, we will present simulations based on including
radiative cooling along with a simple recipe for star formation, but
excluding any form of energy feedback from supernovae. Only in one
case, in which we want to study the structure and the evolution of the
phase diagrams, we turned star formation off. The star
formation recipe adopted assumes that a collisional gas particle, whose
equivalent hydrogen number density exceeds $n_H=0.1$ cm$^{-3}$ and
with temperature below $3\times 10^4$K, is instantaneously converted
into a collisionless ``star'' particle. The practical advantage of
including star formation is that the simulations becomes
computationally much faster, since one avoids performing intensive SPH
computations among cooled high-density gas particles. As we shall
discuss in the Section 4, we verified that the evolution of the mass
deposition rate is essentially independent of the introduction of star
formation.

Initial conditions for isolated halos have been created by placing gas
in hydrostatic equilibrium within a DM halo, having the NFW density
profile (see eq. \ref{nfw}), according to the model given in
section~\ref{section:profile}.  To fix the gas thermal energy, we
required, as suggested by \cite{Komatsu01}, that the slopes of the DM
and gas density profiles be equal at the virial radius.  This leads to
thermal energies very similar to 1.2 times the virial energy, as used
by \cite{Monaco07}.  In order to generate initial conditions, initial
positions of DM and gas particles are generated by Monte Carlo sampling
the analytical profiles of eqs. (\ref{nfw}) and
(\ref{eq:hydro_sol}). To create an equilibrium configuration for the
DM halo, initial velocities of the particles are assigned according to
a local Maxwellian approximation \citep{Hernquist93}, where the width
of the distribution is given by the velocity dispersion of the DM
particles, as obtained by solving the Jeans' equation. As for the gas
particles, their internal energy is assigned according to the third of
the equations (\ref{eq:hydro_sol}). Since the above equations are a
hydrostatic solution, gas particles are initially assigned zero
velocities.

\begin{figure*}
\centerline{
\psfig{figure=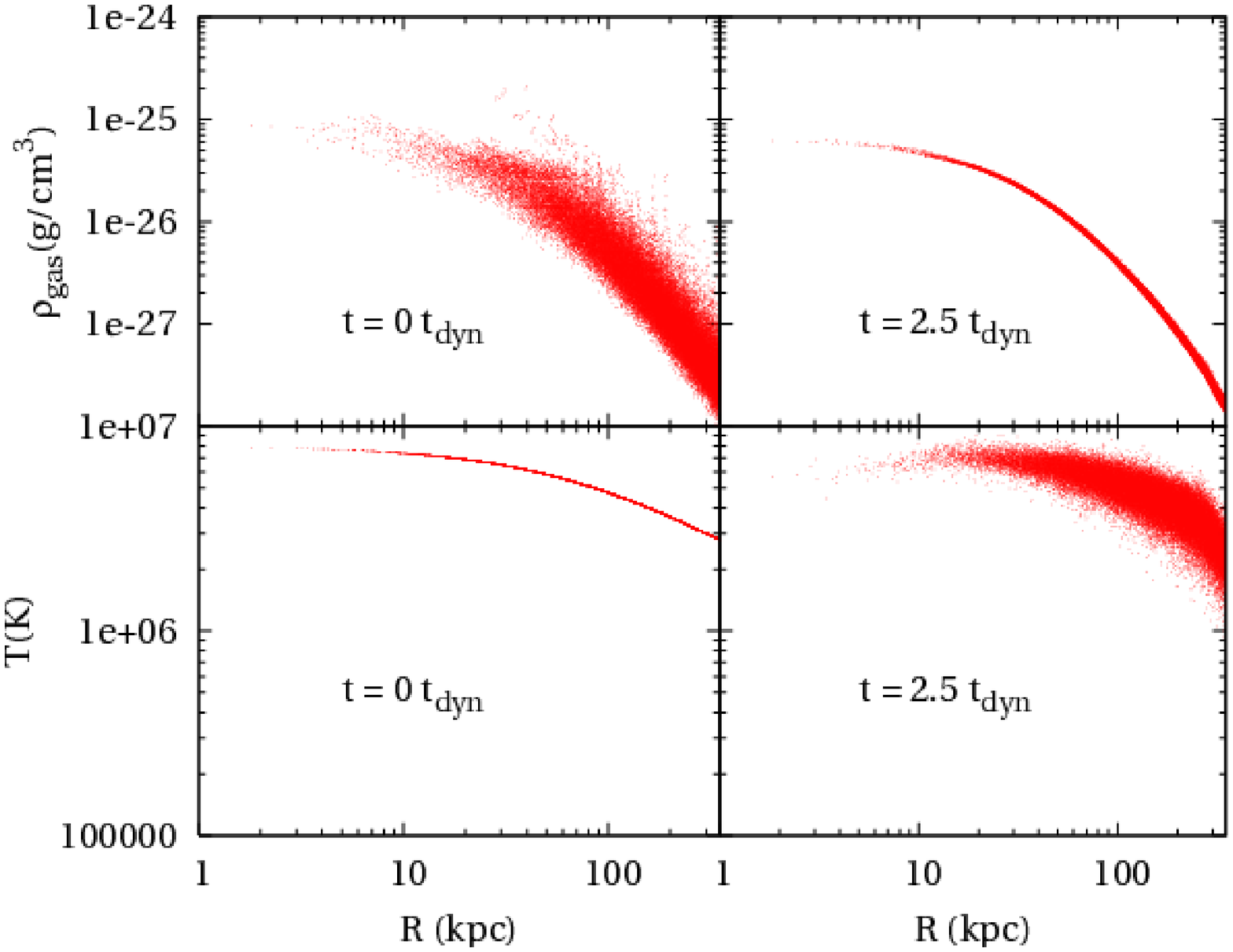,width=12cm,angle=270}}
\caption{\emph{Top panels}: Density of gas particles for the
  non-radiative run of the halo H1, as a function of their
  halo-centric distance, at $t=0$, as assigned in the initial
  conditions (left) and after $2.5 t_{\rm dyn}$ (right).  \emph{Bottom
  panels}: the same as for the top panels, but for the temperature of
  gas particles.}
\label{fig:relax}
\end{figure*}

\begin{table}
\centering
\caption{Characteristics of the simulated halos. Column 1: halo name;
  column 2: halo mass enclosed within $r_{200}$; Column 3: NFW
  concentration parameter; Column 4: redshift used to compute the
  reference critical density; Column 5: value of $r_{200}$ (in kpc);
  Columns 6: dynamical time (in Gyr); Column 7: central cooling time
  (in Gyr).}
\begin{tabular}{|r|r|r|r|r|r|c|}
\hline
$$&$M_{200}$&$c_{\rm nfw}$&redshift&$r_{200} $&$t_{\rm dyn}$&$t_{\rm cool,0}$\\
\hline
$H1$&$10^{13}$&$6.3$&$z=0$&$350$&$0.56$&$0.56$\\
$H2$&$10^{13}$&$7.25$&$z=0$&$350$&$0.56$&$0.40$\\
$H3$&$10^{13}$&$5.25$&$z=0$&$350$&$0.56$&$0.82$\\
$H4$&$10^{12}$&$7.25$&$z=0$&$162$&$0.56$&$0.12$\\
$H5$&$10^{15}$&$4.3$&$z=0$&$1623$&$0.56$&$6.67$\\
$H6$&$3\cdot10^{11}$&$6.53$&$z=1$&$75$&$0.32$&$0.04$\\
$H7$&$10^{13}$&$5.63$&$z=1$&$241$&$0.33$&$0.31$\\
$H8$&$10^{12}$&$5.92$&$z=2$&$79$&$0.19$&$0.04$\\
\hline
\end{tabular} 
\end{table}

Each halo has been sampled with $6\times 10^4$ DM particles inside
$r_{200}$ and an initially equal number of gas particles. The ratio
between the mass of DM and gas particles is determined by the baryon
fraction, that we assume to be $f_{\rm bar}=0.19$. As for the choice of
the gravitational softening, it has been chosen to be about
three times larger than the lower limit recommended by \cite{Power03},
$\epsilon \simeq 3r_{200}/\sqrt{N_{200}}$ for a Plummer-equivalent
softening, where $N_{200}$ is the number of particles within
$r_{200}$. The minimum value allowed for the SPH smoothing length is
assumed to be 0.5 times the value of the gravitational softening, 
with SPH computations performed using $N_{\rm ngb}=32$ for the number of
neighbours.

To ensure stability of the halos in the absence of cooling, density
profiles have been sampled with particles out to about
$8r_{200}$. This also ensures an adequate reservoir of external gas
than can flow in while cooling removes pressure support in the central
halo regions. In principle, our controlled numerical experiments could
have been performed by using a static NFW potential, instead of
sampling the halo with DM particles. However, this procedure would not
have allowed us to account for any backreaction of cooling on the DM
component. Indeed, it is known from cosmological simulations that
including gas cooling causes a sizeable change (adiabatic
contraction) of the structure of the DM halos \citep[e.g.,
][]{Kravtsov04}.

The characteristics of the simulated halos are summarized in Table
1. In this table we also specify the redshift at which the simulated
halo is assumed to stay. Although redshift never explicitly enters in
our simulations of isolated halos, it appears in a indirect way when
we fix the value of the critical density. Ultimately, increasing
redshift amounts to take a higher value of $\rho_{\rm crit}$ and,
therefore, a higher halo density for a fixed value of $M_{200}$. In
this way, we expect that ``high-redshift'' halos will have a shorter
cooling time. In the following, we assume the relation between
redshift and critical density to be that of a cosmological model with
$\Omega_m=0.3$ and $\Omega_\Lambda=0.7$.

Our reference halo has a mass $M_{200}=10^{13}M_\odot$ (H1 in Table
1), typical of a poor galaxy group, with a value of the concentration
parameter $c_{\rm nfw}=6.3$, given by the relation between mass and
concentration provided by NFW, and $r_{200}$ corresponding to the
value of $\rho_{\rm crit}$ at $z=0$. Two other halos of the same mass are
also simulated, which have different values of the concentration
parameter, still lying within the scatter in the $M_{200}$--$c_{\rm nfw}$
relation (H2 and H3). We then simulate a smaller (H4) and a larger
(H5) halo, with $M_{200}=10^{12}M_\odot$ and $M_{200}=10^{15}M_\odot$,
respectively, so as to sample also the scale of elliptical galaxies
and of rich clusters. We finally consider two halos at $z=1$ (H6 and
H7) and one halo at $z=2$ (H8). In all runs we used the same value,
$\gamma_p=1.18$, for the effective polytropic index.  In Table 1 we
also report for each halo the values of the dynamical time-scale,
which is defined as
\begin{equation}\label{tdyn}
t_{\rm dyn}=\Bigg(\frac{1}{4\pi G \rho}\Bigg)^{1/2}\,,
\end{equation}
and that of the cooling time calculated at the center of the halo
as in equation (\ref{eq:coolingtime}).

In order to verify the numerical stability of our results, we also
performed the following runs for the H1 halo: {\em (a)} a simulation
with a 10 times larger number of particles and a rescaling of the
softening, in order to check the effect of mass resolution (H1-HR);
{\em (b)} a simulation with a 4 times larger number of particles
(at fixed halo mass) {\em and} number of
neighbours $N_{\rm ngb}$ than in the reference run, keeping the
softening constant for the gravitational force (H1-4SPH): because
mass resolution is given by $N_{\rm ngb}$ times particle mass, this is
kept constant while decreasing the discreteness noise in the SPH
computation; {\em (c)} a simulation with cooling only and without star
formation (H1-C); {\em (d)} a simulation in which the gravitational
softening is halved with respect to the reference value (H1-S).

\begin{figure}
\centerline{
    \psfig{figure=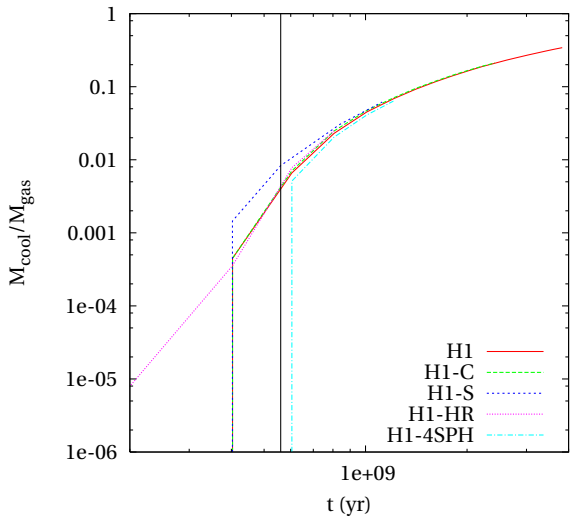,width=7.0cm,angle=270}}
  \caption{Evolution of the cooled mass $M_{\rm cool}$ for different runs
    of the H1 halo. Solid line: reference run with cooling and star
    formation; dashed curve: run with only cooling (H1-C);
    short-dashed curve: run with cooling and star formation, with
    halved gravitational softening (H1-S); dotted curve: run with
    cooling and star formation at 10 times better mass resolution
    (H1-HR); dotted-dashed curve: run with cooling and star formation
    at 4 times more particles and SPH neighbours (H1-4SPH) . The
    vertical black line represents the theoretical central cooling
    time.}\label{fig:stable}
\end{figure}

All the initial conditions have been first evolved for 10 dynamical
times, without cooling. This allowed us to check that temperature and
density profiles are always stable, thus confirming that the initial
conditions are indeed quite close to configurations of hydrostatic
equilibrium. An example of this is shown in Figure
\ref{fig:relaxed_profile}, where we plot the profiles of gas density,
DM density and temperature for the reference halo H1, at different
epochs. Although the profiles are all remarkably stable, it is worth
reminding that there are at least two reasons why our initial
conditions may not be equilibrium configurations.  Firstly, the gas
profiles given by equations (\ref{eq:hydro_sol}) are not an exact
equilibrium solution because the gas mass is not negligible. Secondly,
initial particle positions are assigned by performing a Monte Carlo
sampling of the gas and DM density profiles, while the internal energy
of gas particles is assigned in a deterministic way, by using the
third of the equations (\ref{eq:hydro_sol}). The scatter associated to
the Poissonian sampling of the density profiles, joined with the
deterministic assignment of the internal energy of SPH particles, may
well not represent a configuration of stable equilibrium. If this is
the case, we then expect that the system relaxes to the true minimum
energy configuration in a time comparable to its dynamical
time-scale.  Indeed, Figure~\ref{fig:relax} shows the scatter plot of
individual values of density and temperature of all the gas particles,
as a function of their halo-centric distance, for the initial
conditions of H1 (left panels) and for a configuration obtained by
evolving the system for $2.5 t_{\rm dyn}$ in the absence of cooling
(right panels). The relaxed configuration shows residual scatter
both in density (amounting to 5 per cent) and temperature (amounting
to 15 per cent).  The same amount of scatter is also found in the high
resolution run (H1-HR), while a smaller amount (3 per cent in
density, 5 per cent in temperature) is found in the H1-4SPH
simulation, where the density is computed averaging over 128 instead
of 32 particles. The latter result suggests a numerical origin for
this scatter, related to the finite number of particles used in the
SPH computations.

In order to account for this relaxation, we decided to use as initial
conditions for the radiative runs the configuration attained by
evolving the non-radiative runs for $2.5$ dynamical times. Once
cooling is turned on, all simulations are let then evolving for 8
central cooling times, with the exception of H5, which is run for roughly
two central cooling times.

\section{Results}
As already emphasized, the main aim of our analysis is to understand
the radiative cooling of the gas in DM halos and to point out which
one of the cooling models, described in section 2, gives results on
the evolution of the cooled gas mass which are in best agreement with
the numerical experiment. To this purpose, most of the discussion on
how gas cools in simulations will refer to the H1 halo. Since the
evolution of $M_{\rm cool}$ is the central result of our analysis, we will
show it for all the halos and compare the simulation results with the
predictions of the cooling models.

In Figure \ref{fig:stable} we show the evolution of $M_{\rm cool}$ for the
reference run of the halo H1 (i.e. cooling and star formation), and
compare it with the same run without star formation (H1-C), with the
run having 10 times better mass resolution (H1-HR), four times more
particles and SPH neighbours (H1-4SPH) and with the run with standard
mass resolution but with gravitational softening smaller by a factor
of two (H1-S). In the runs with cooling and star formation, $M_{\rm cool}$
is contributed both by the mass in collisionless stars and by the mass
in cold gas particles, which have temperature below $3\times 10^4$
K. In the run with cooling only, $M_{\rm cool}$ is clearly contributed
only by particles colder than the above temperature limit. Figure
\ref{fig:stable} shows that the evolution of the cooled mass is
independent of whether cold and dense particles are treated as
collisionless or SPH particles. This result, which agrees with that
found by \cite{Tornatore03} for cosmological simulations of clusters,
confirms that using the SPH scheme with explicit entropy conservation
and arithmetic symmetrization of hydrodynamical forces
\cite{SprHern02} is able to suppress the spurious gas
cooling which otherwise takes place at the interface between cold and
hot phases. This result also demonstrates that including star
formation in the radiative runs does not affect our results on
$M_{\rm cool}$. As for the run with a larger number of neighbours
(H1-4SPH), cooling turns out to start at a later epoch. The reason for
this lies in the reduced scatter in density in the initial conditions,
when a larger number of neighbours for the SPH computations is
used. In fact, a gas particle whose density is scattered upwards with
respect to the density computed from the profile at its halo-centric
distance, has a shorter cooling time. As a consequence, the smaller the
scatter, the lower the probability that a gas particle has cooling
time significantly shorter than that relative to its radial
coordinate.  As for the high-resolution run (H1-HR), the larger
number of particles provides a better sampling of the scattered
density distribution. Therefore, there is an increasing
probability to have a small number of gas particles, with
exceptionally up-scattered density, which cool down at earlier
times. Finally, decreasing the gravitational softening by a factor 2
(H1-S run) leads to better resolving the very central part of the
halo, where gas can more easily cool. This turns into
a stronger initial transient in the evolution of $M_{\rm cool}$ at the
onset of cooling. Quite reassuringly, despite the differences in the
evolution of $M_{\rm cool}$ between these runs during the first cooling
time, they all nicely converge after about two cooling times. This
demonstrates that our numerical description of
the evolution of the cooled mass is numerically stable after an
initial transient.

\begin{figure*}
\centerline{
    \psfig{figure=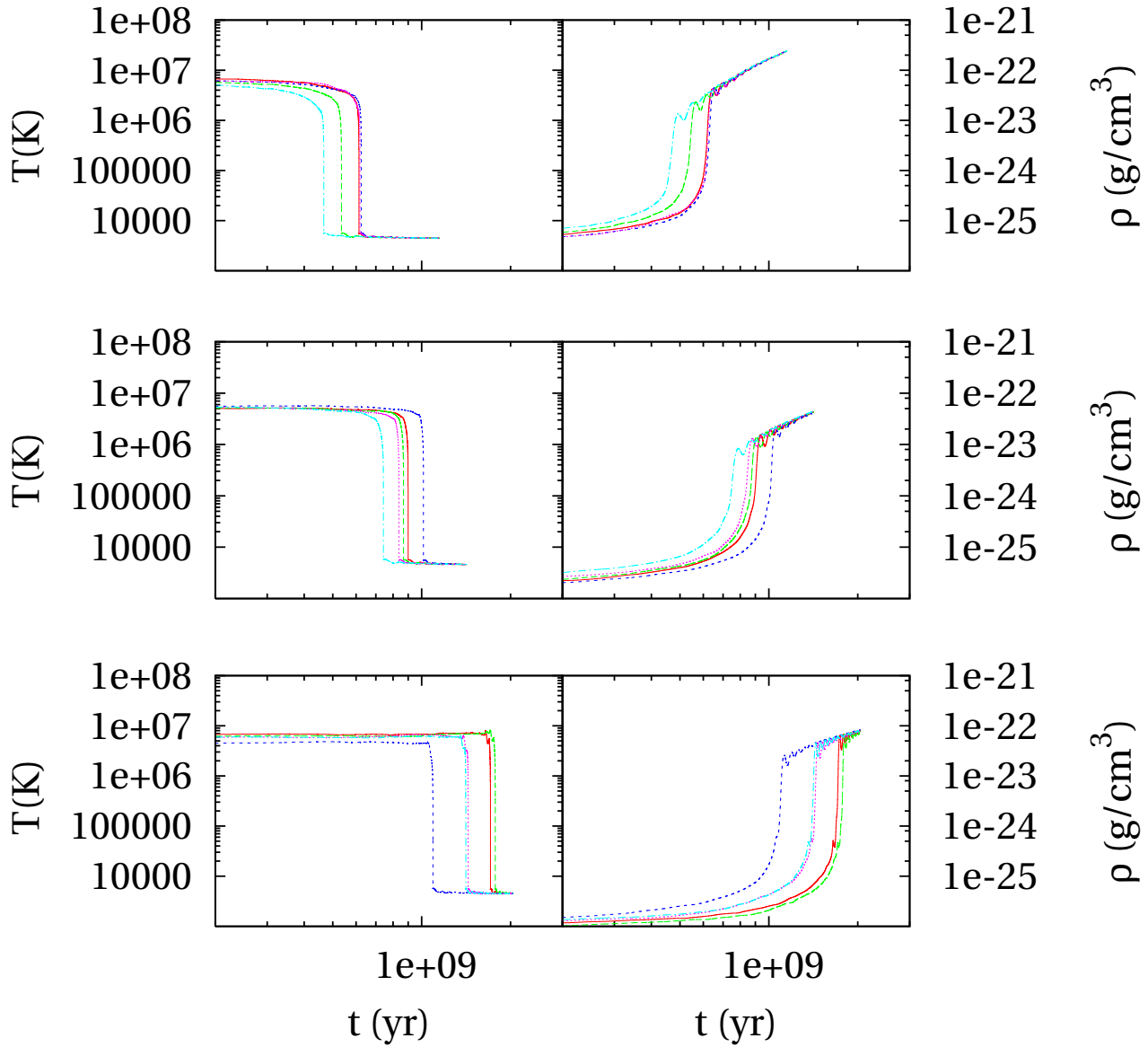,width=18cm,angle=270}}
  \caption{Evolution of temperature (\em{left panels}) and density ({\em right
    panels}) for three sets of five particles each, selected within
    three different radial shells in the initial conditions. {\em Top
    panels}: $(r_{\rm min},r_{\rm max})=(10,20)$ kpc; {\em Central
    panels}:$(r_{\rm min},r_{\rm max})=(30,40)$ kpc. {\em Bottom panels}:
    $(r_{\rm min},r_{\rm max})=(50,60)$ kpc.}\label{fig:temptime}
\end{figure*}

In order to probe in detail the behaviour of gas in radiative
simulations, we have randomly selected five particles in three
distance intervals from the center (10-20 kpc, 30-40 kpc and 50-60
kpc) in the initial conditions and we have followed their evolution in
the H1-C run (the absence of star formation allows to follow the
transition from the hot to the cold phase). In Figure
\ref{fig:temptime} we plot the evolution of temperature and density as
a function of time for the selected particles.
While flowing toward the halo centre, each particle roughly maintains
its initial temperature while its density progressively
increases. Afterwards, in a very short time interval, it cools down to
$\mincir 10^4$K, which corresponds to the temperature where the
cooling function dies. This means that the transition from the hot to
the cold phase takes place quite rapidly, thus keeping the two phases
well separated.

\begin{figure*}
\begin{center}
\hbox{
\psfig{figure=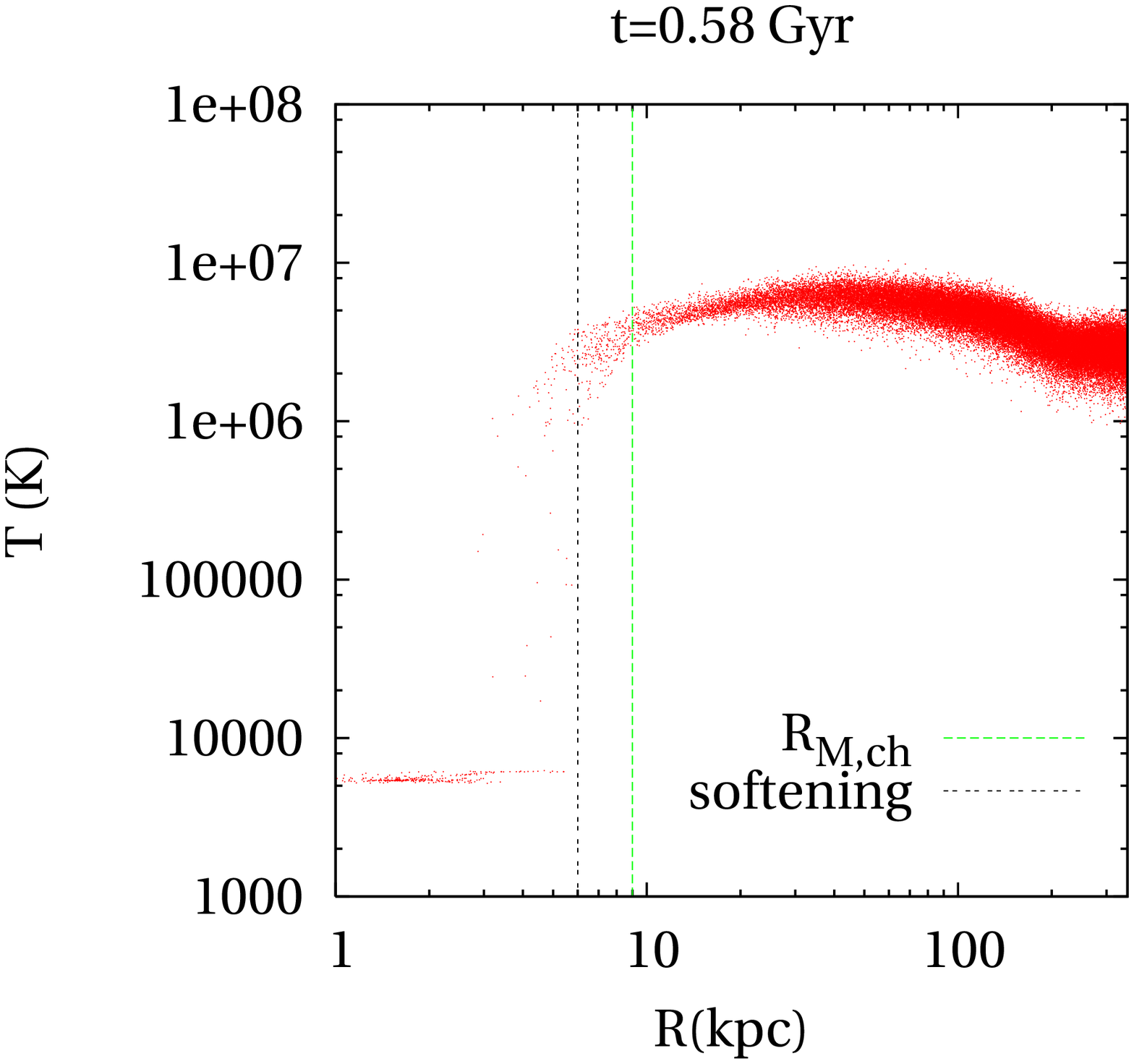,width=8cm,angle=270}
\psfig{figure=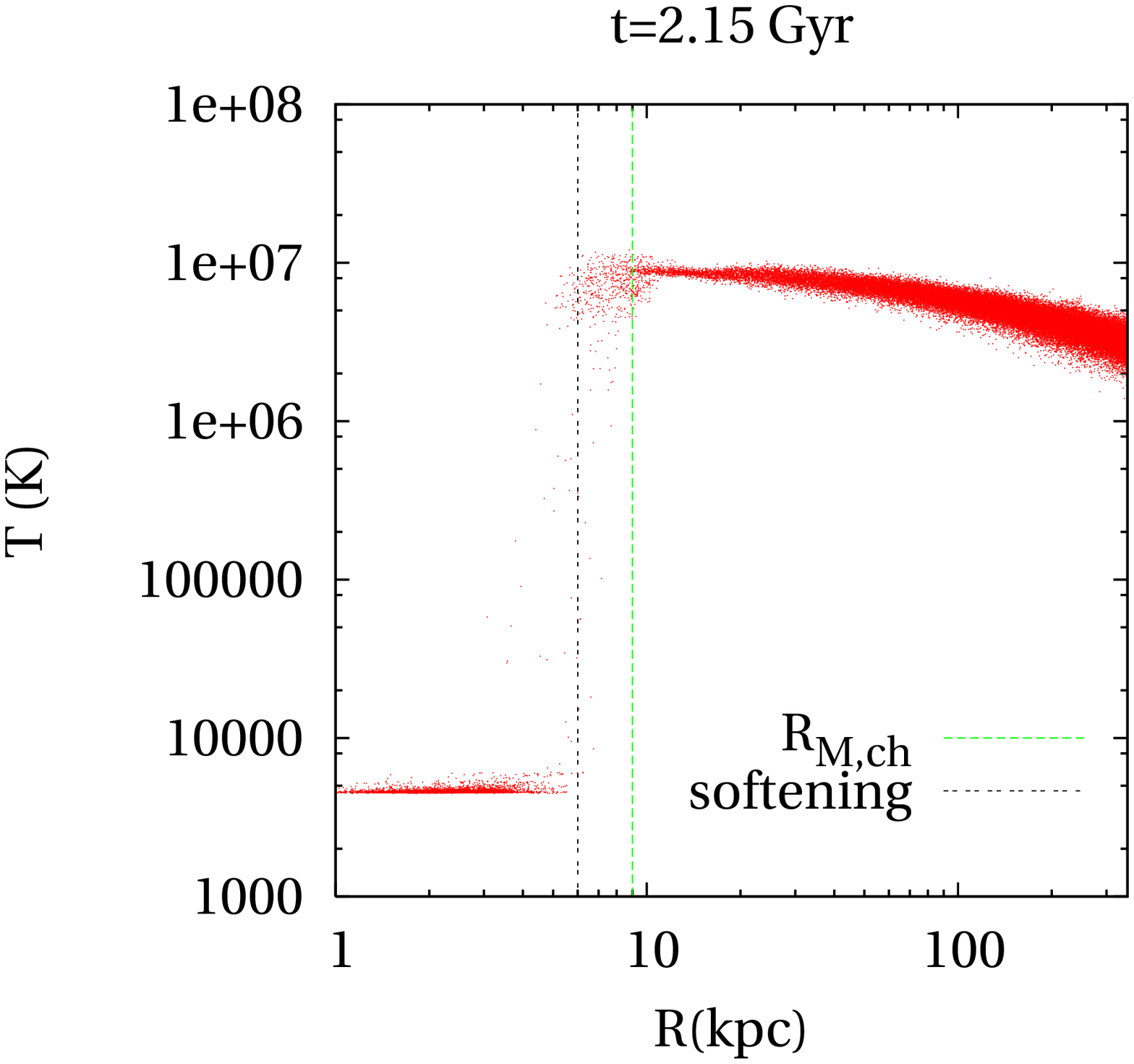,width=8cm,angle=270}
}
\end{center}
\caption{Scatter-plot of temperature vs. halo-centric distance for gas
  particles in the run of the H1 halo which includes only cooling,
  without star formation (H1-C). The green, dotted lines give the
  cooling radius $r_{\rm M,ch}$ as predicted by the closed {\sc
  morgana} model, while the black, dashed lines denote the softening
  scale.  }
\label{fig:eul_temp}
\end{figure*}

Figure~\ref{fig:eul_temp} shows the scatter plot of temperature
vs. halo-centric distance of the gas particles in the H1-C run at two
output times.  It is possible to identify a ``cooling region'' as the
spherical shell where the drop in temperature takes place. It is
interesting to note that the size of this region is roughly constant
in time and comparable to the softening scale.  We have verified that
the presence of a sharp physical boundary separating hot and cold gas
phases is robust against numerical resolution, in fact with an even
sharper boundary at higher resolution, but its size does depend on
resolution. In the H1-HR run, who has roughly a two times smaller
softening, the size of the cooling region is more than 50 per cent
smaller. Therefore, while our simulations provide a numerically
convergent result on the evolution of $M_{\rm cool}$, they do not
provide a similarly convergent result on the size of the cooling
region.

\begin{figure*}
\centerline{\hbox{
\psfig{figure=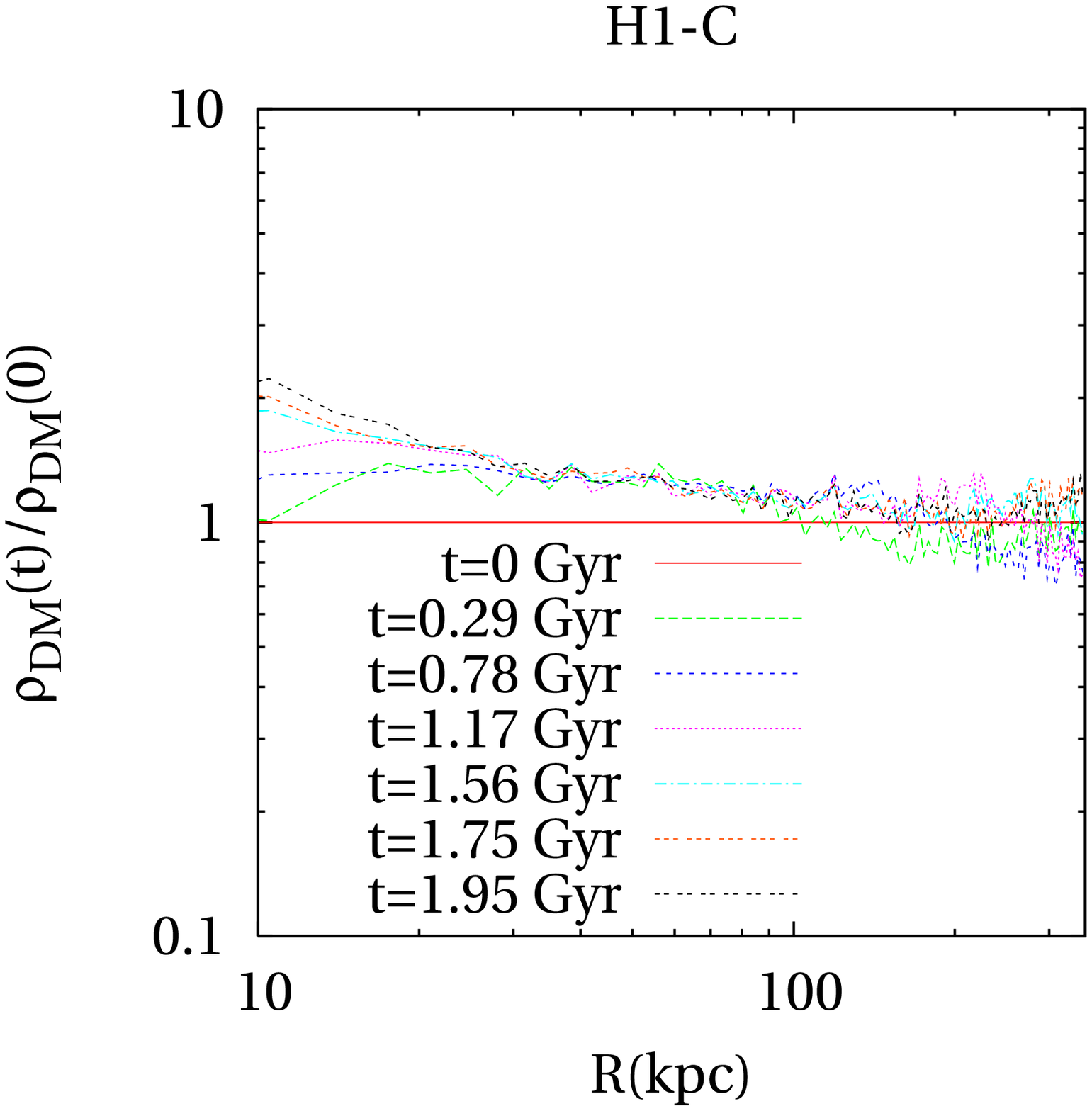,width=8.0cm,angle=270}
\psfig{figure=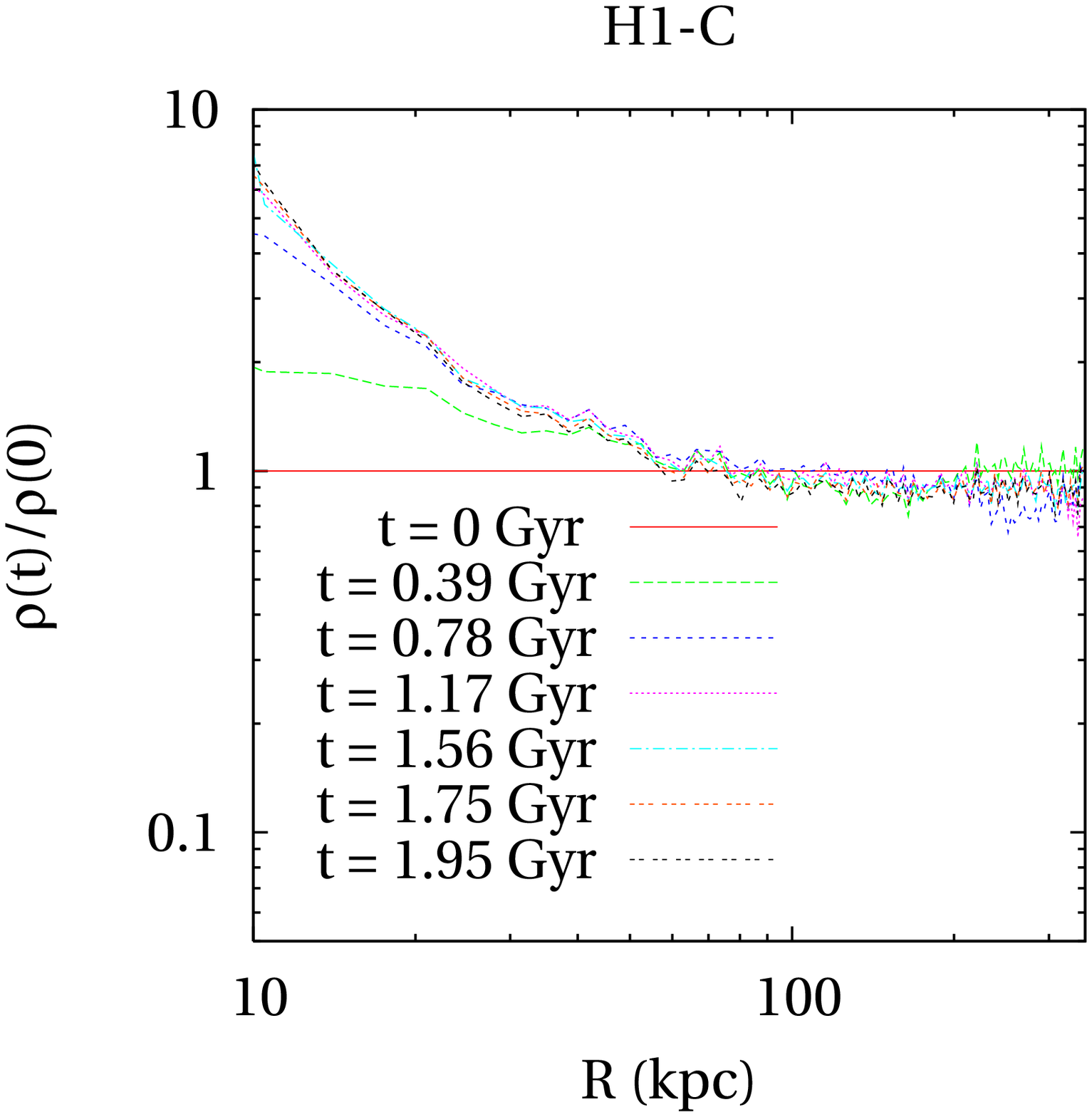,width=8.0cm,angle=270}}}
\centerline{\hbox{
\psfig{figure=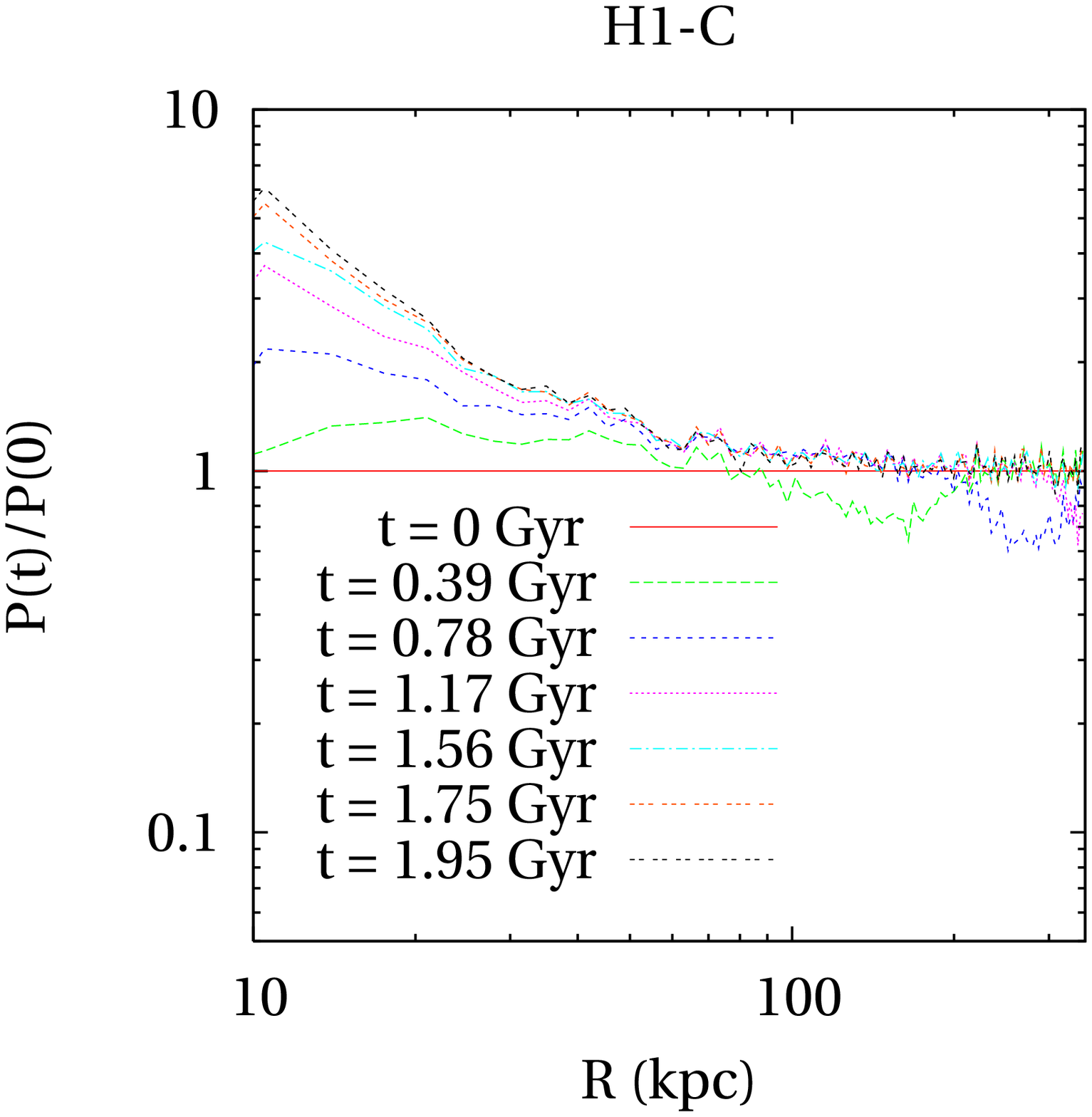,width=8.0cm,angle=270}
\psfig{figure=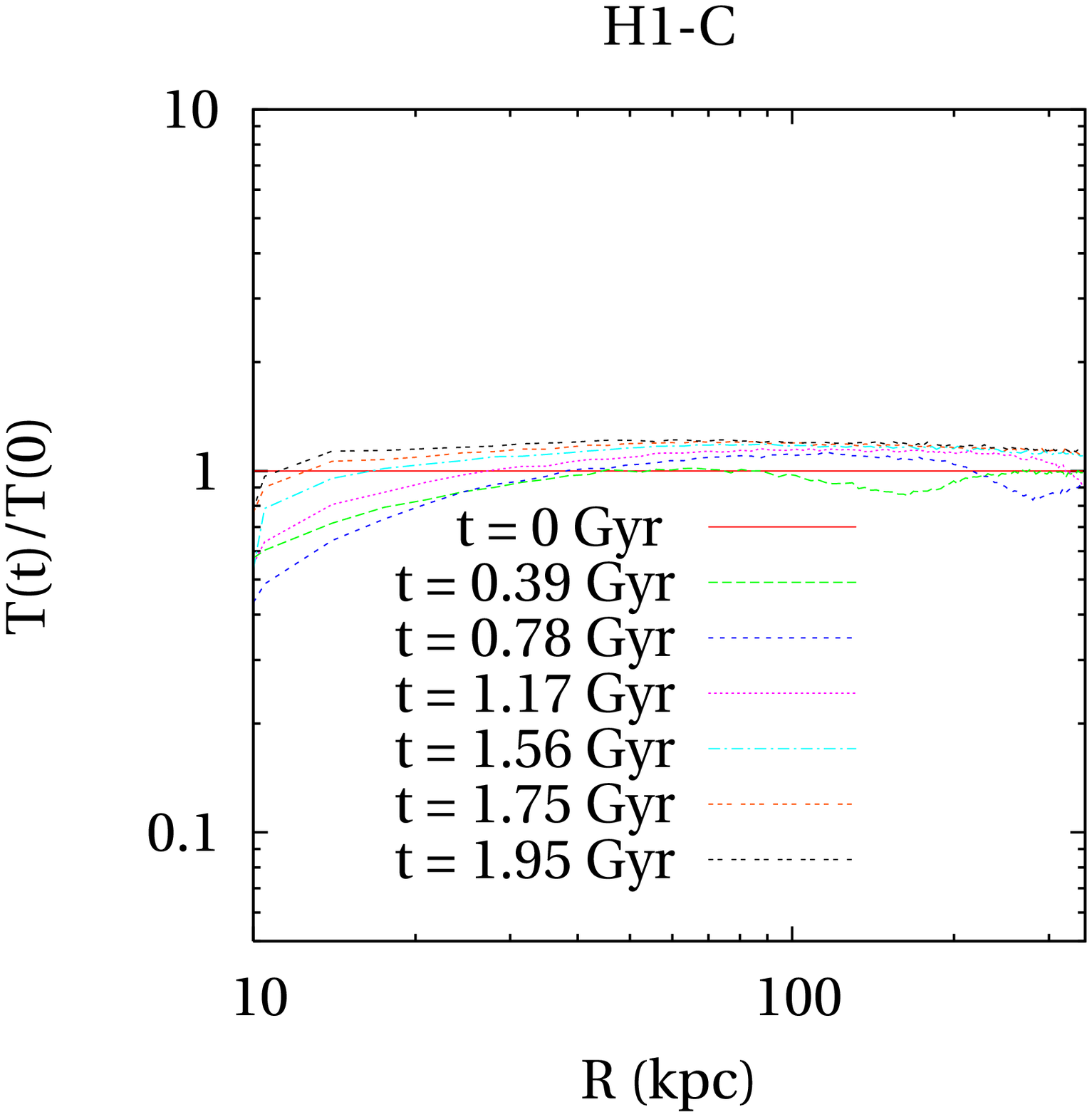,width=8.0cm,angle=270}}}
  \caption{Profiles of DM density (upper left panel), gas density
(upper right), pressure (lower left) and temperature (lower right) at
different epochs for the H1 simulation with only cooling (H1-C).  All
profiles are normalized to their initial value.}\label{fig:pressure}
\end{figure*}

In Figure~\ref{fig:pressure} we show the density profile of DM
and the density, pressure and temperature profiles of the hot gas in
the simulation at several times, normalized to the corresponding
profiles evaluated for the initial conditions. Pressure increases in
the inner part of the halo for a few dynamical times, to saturate
later to values that peak at a factor of six times higher than
in the initial conditions, just outside the cooling region. This
increase in pressure is mainly driven by a comparable increase in gas
density, while the temperature profile is much more stable or even
slightly decreasing near the cooling region. The density increase
on the other hand is partly caused by the adiabatic contraction of the
DM halo, but because the increase in DM density, though significant, is
smaller than that in gas density, then adiabatic contraction gives only
a minor contribution.  As we shall discuss in the following, the
increase in gas density enhances radiative losses as the gas
approaches the cooling region, and this turns into a shortening the
cooling time of the inflowing gas particles, with respect to the
predictions of the classical model.

In summary, the following conclusions can be drawn from our analysis
of gas cooling in simulations: \\
{\bf (i)} the drop in temperature when gas
particles pass from the hot to the cold phase is quite rapid;\\ {\bf (ii)}
this drop in temperature takes place in a spherical shell with a
rather sharp boundary, the cooling region, which separates the inner
and outer regions dominated by cold and hot particles, respectively;\\
{\bf (iii)} density and pressure increase with time just beyond the
cooling region, 
and this increase is not driven by adiabatic contraction of the halo.

In the light of these results, the question then arises as to whether
the cooling recipes, described in section~2, able to reproduce the
rate of mass cooling found in the simulations.

In order to answer this question, we first address the issue
concerning the tuning of the initial conditions. As mentioned in
section 3, the gas in the simulations relaxes to a minimum energy
configuration, so that the parameters of the gas profile after 2.5
dynamical times, when the cooling is turned on, may in principle
differ from the ones used to generate the initial conditions. This is
a crucial point because in order to make a reliable comparison between
analytic cooling models and numerical simulations, we must be sure
that the initial conditions are the same for both. Owing to the
stability of the profiles shown in Figure \ref{fig:relaxed_profile},
we expect this effect to be small. As we shall see, even small
differences in the initial profiles leads to an appreciable change in
the resulting evolution of $M_{\rm cool}$.  In the model that we used
to generate the initial hydrostatic configurations
(equations~\ref{eq:hydro_sol}), the parameters that determine the gas
density and the temperature profiles are the halo gas mass $M_g$, the
effective polytropic index $\gamma_p$, and the central temperature of
the gas (in units of the virial one) $\eta$. While holding the mass
fixed, we have considered 900 pairs of values for the parameters
$(\gamma_p,\eta)$, varying both the polytropic index and the energy
factor in the range $[1.1:1.4]$. We have calculated for each pair the
theoretical density and temperature profiles according to equations
(\ref{eq:hydro_sol}), and compared them with the profiles from the
initial conditions. In particular, we calculated for each radius the
root mean square difference between the density and temperature
profiles of the hydrostatic model and those of the initial conditions,
and imposed the maximum difference to be smaller than 10 per
cent. Then, for each cooling model and unless otherwise stated, we
show predictions relative to all the profiles that were selected by
the procedure described above. This allows us to keep control on any
uncertainty of the initial profiles which are used as input to the
cooling models.

\begin{figure*}
\begin{center}
\centerline{
\vbox{
\hbox{
\psfig{figure=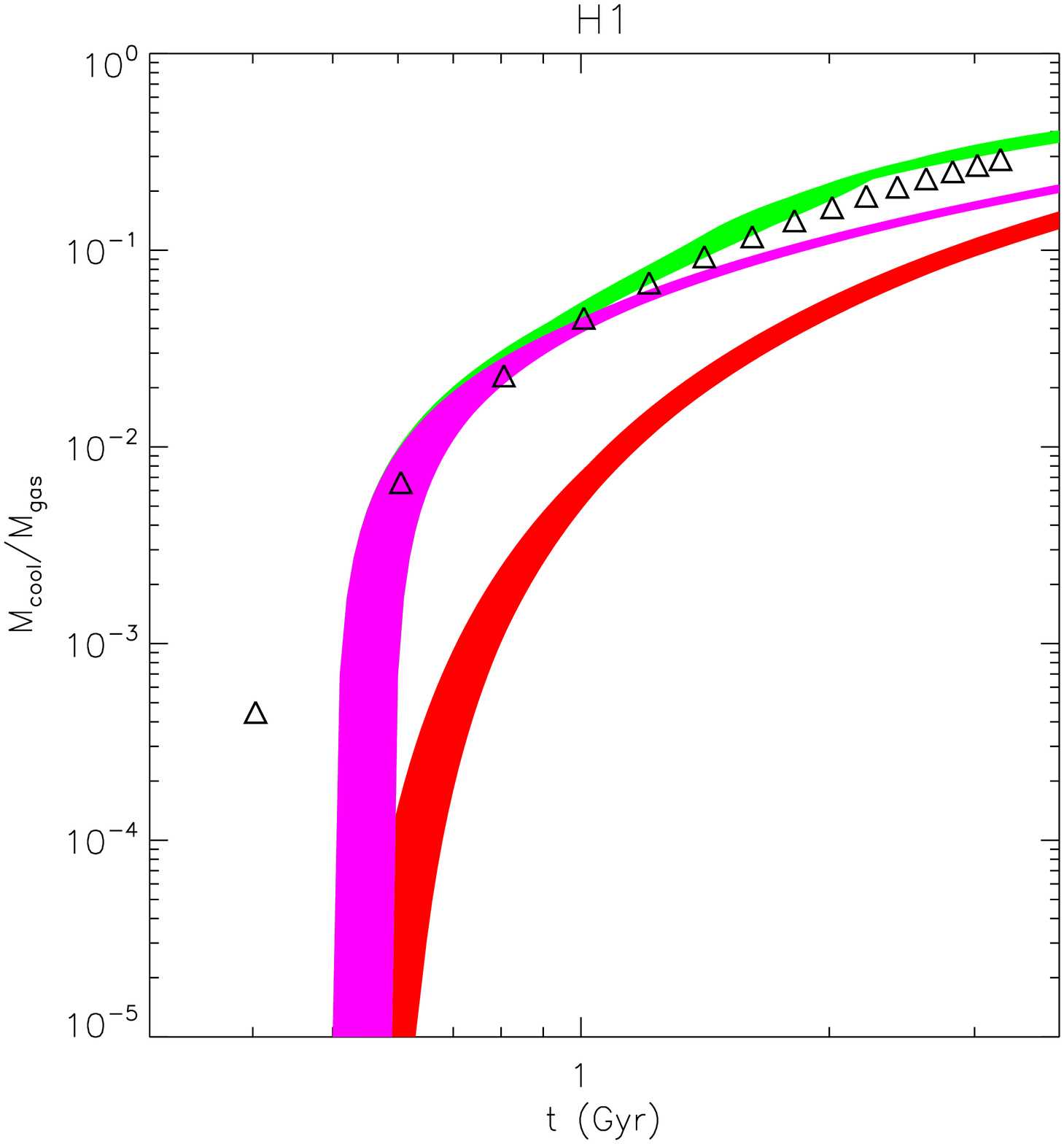,width=6cm}
\psfig{figure=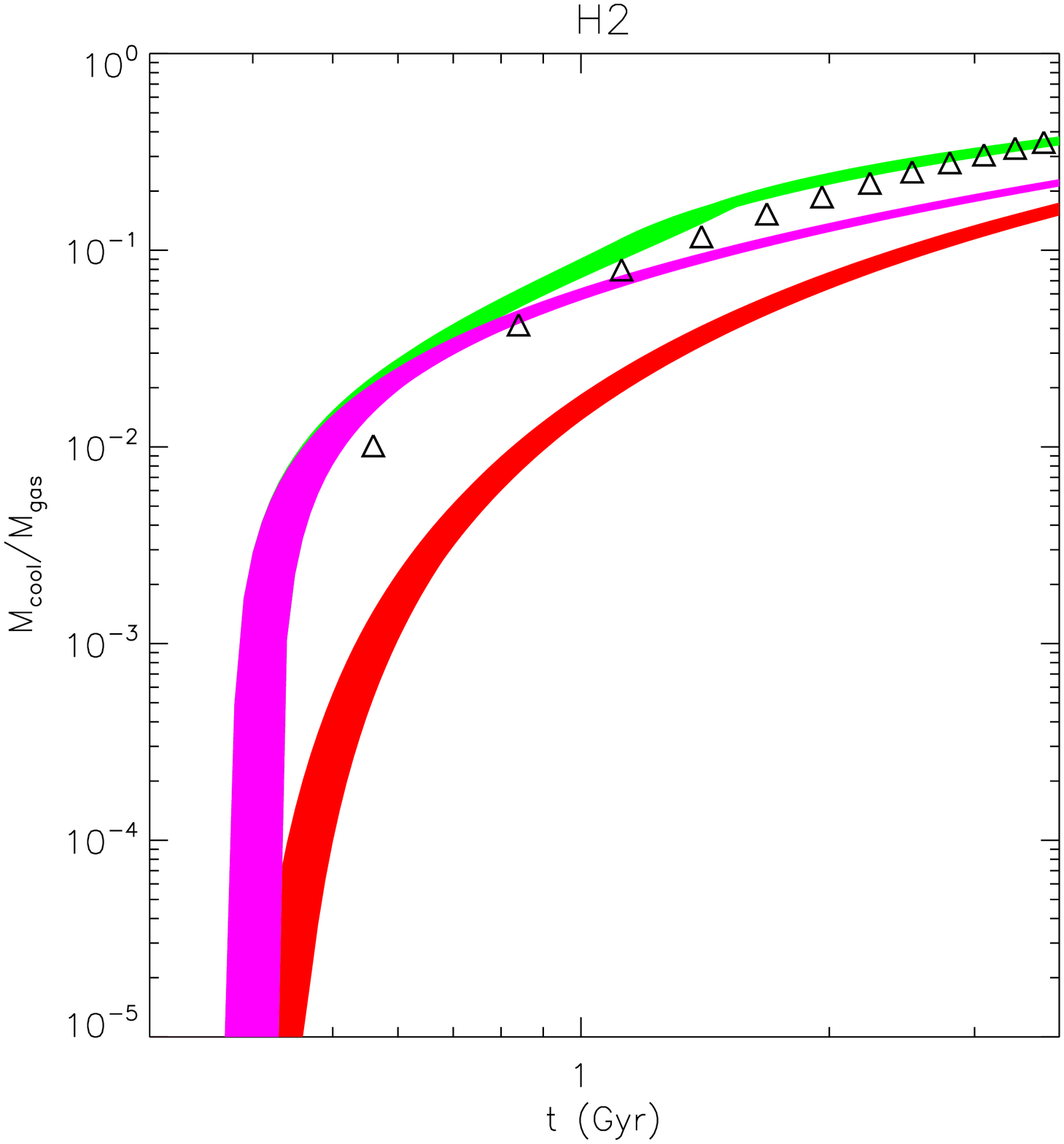,width=6cm}
\psfig{figure=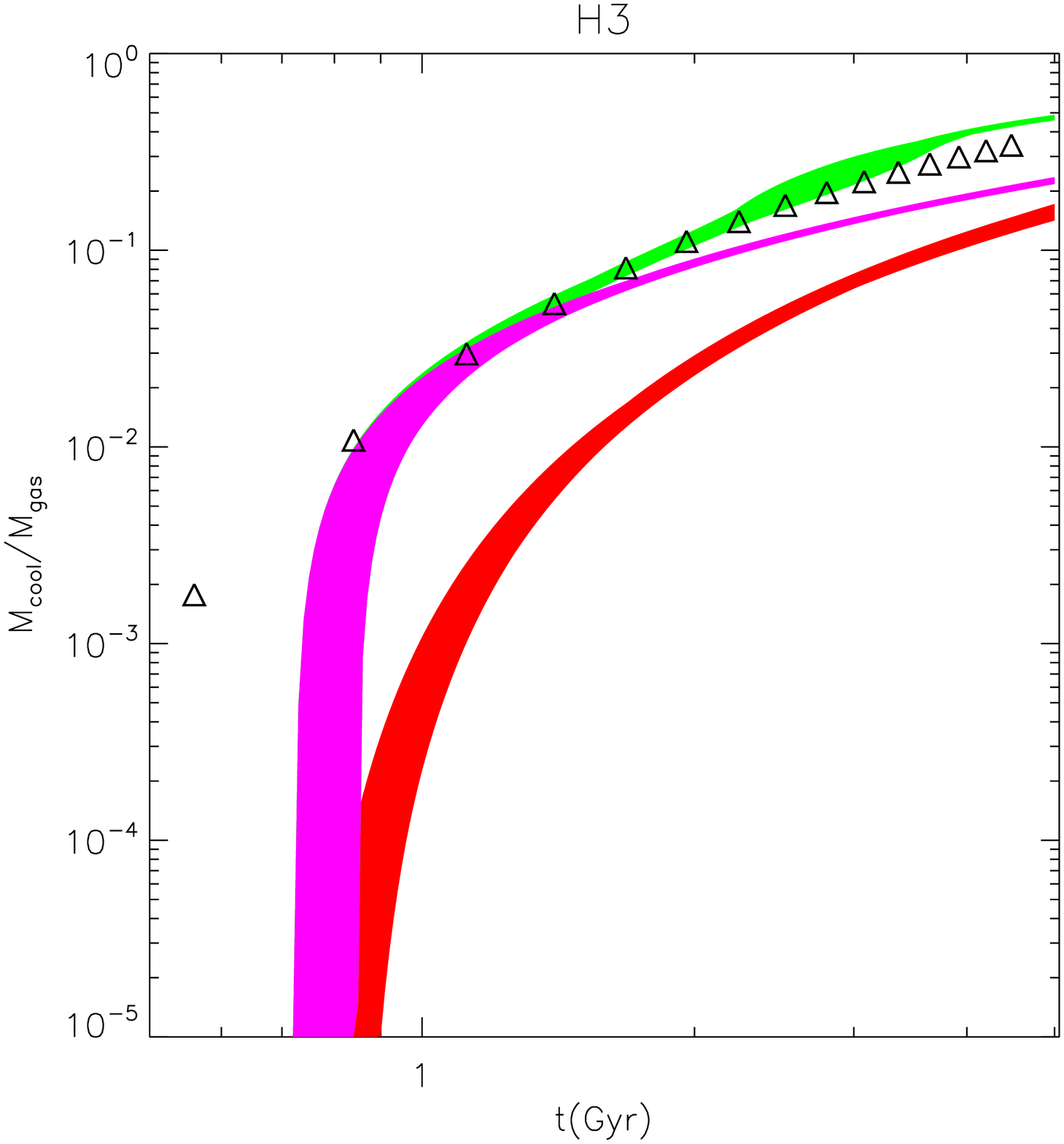,width=6cm}
}
\hbox{
\psfig{figure=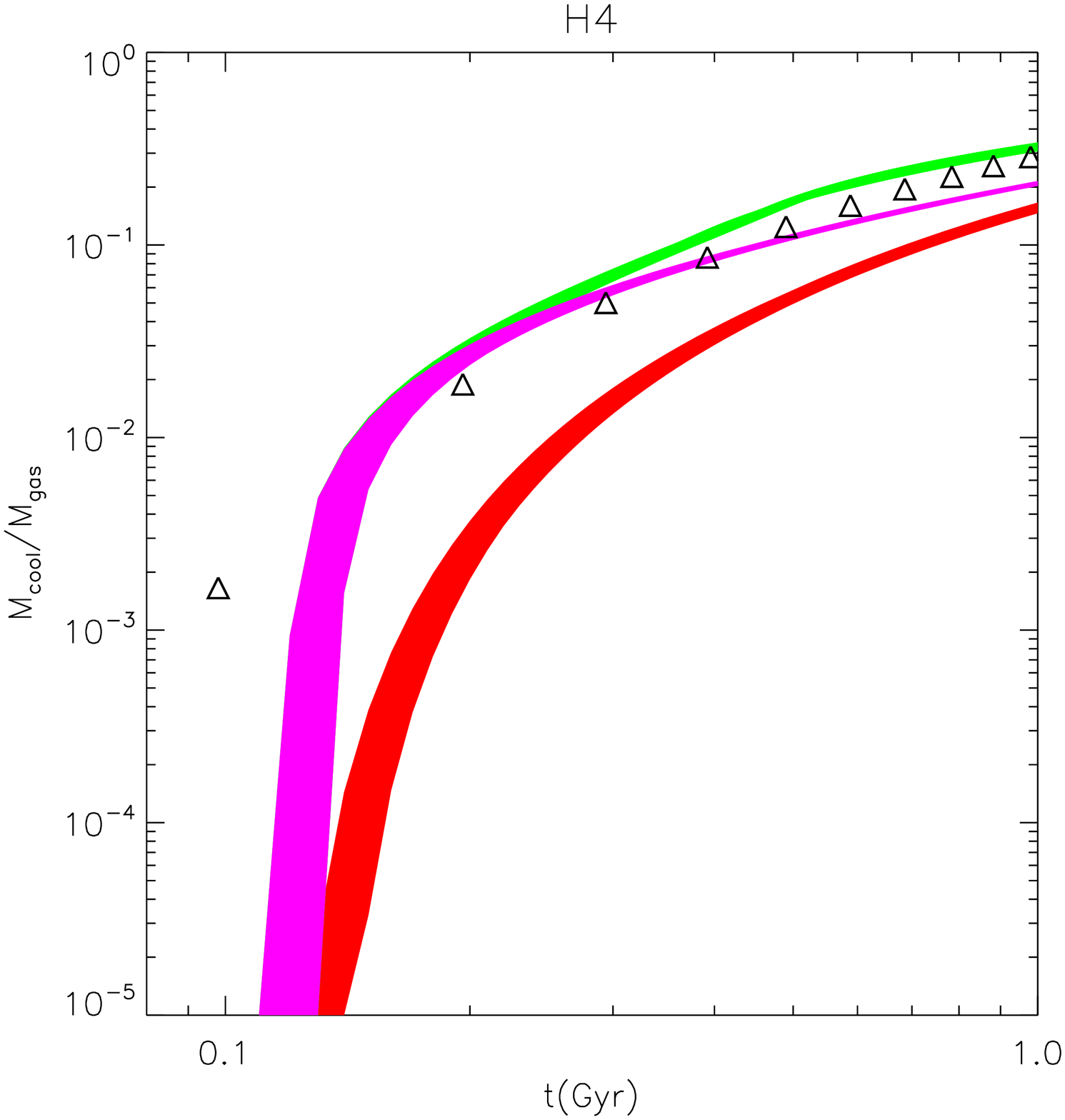,width=6cm}
\psfig{figure=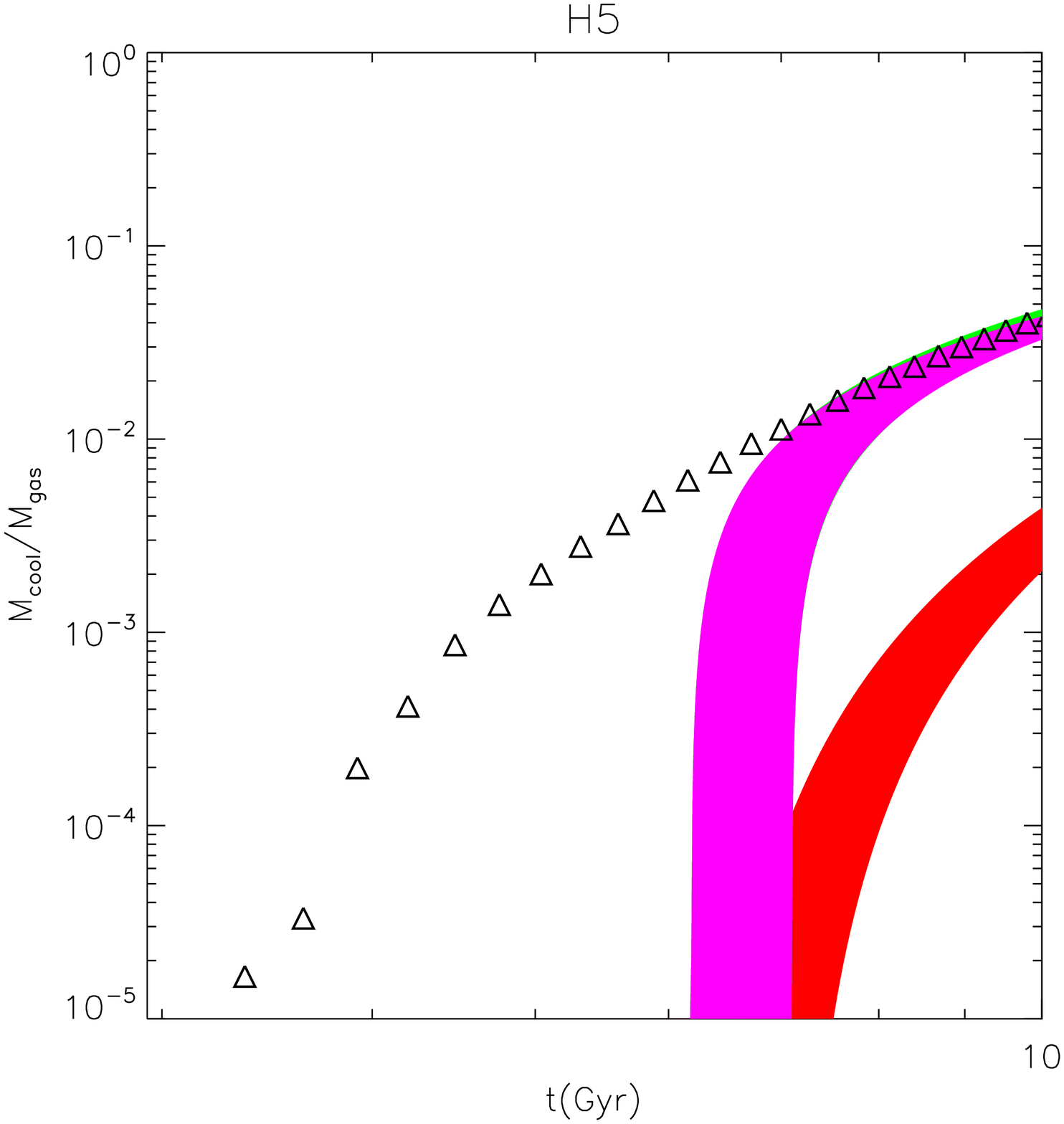,width=6cm}
\psfig{figure=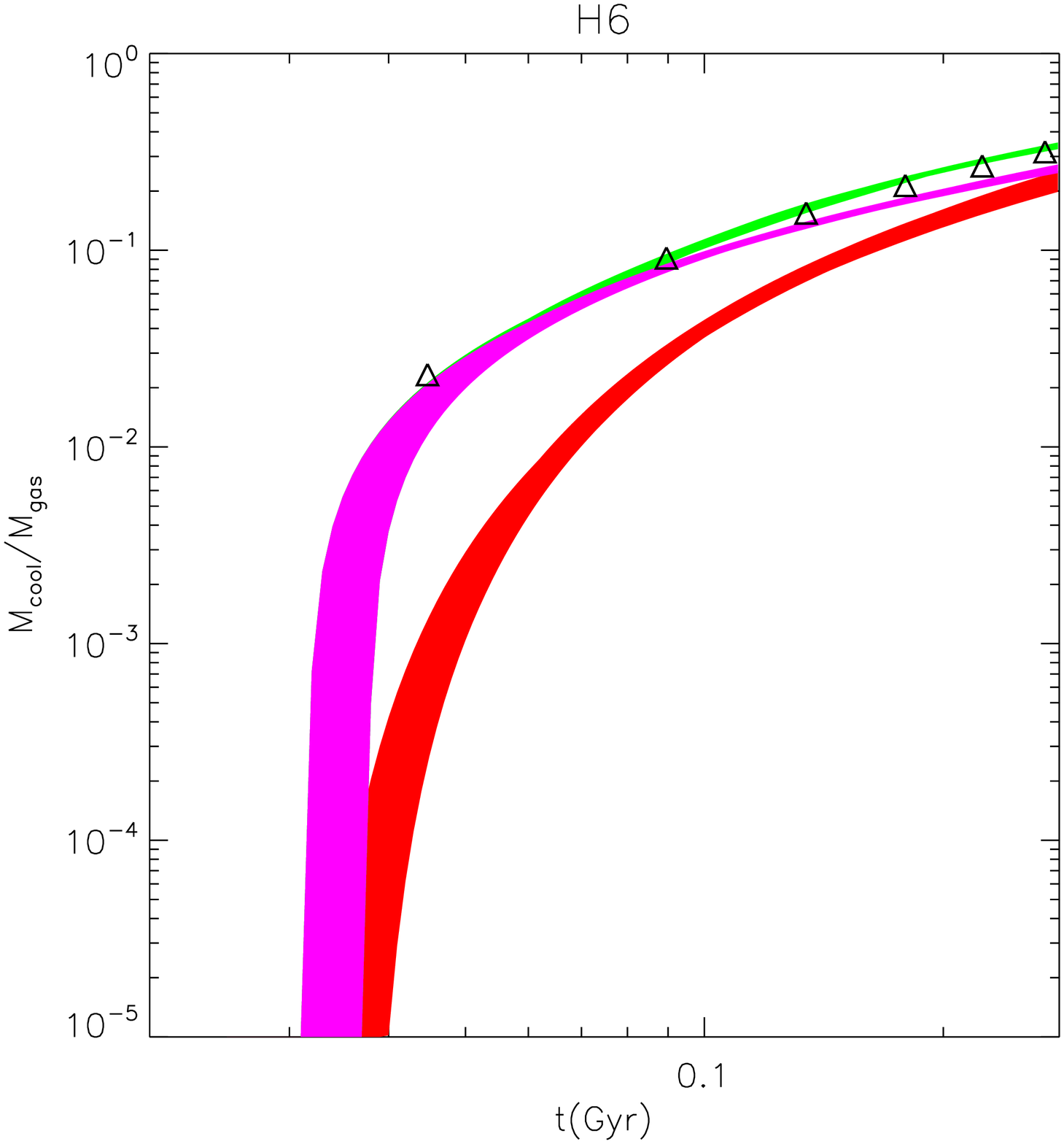,width=6cm}
}
\hbox{
\psfig{figure=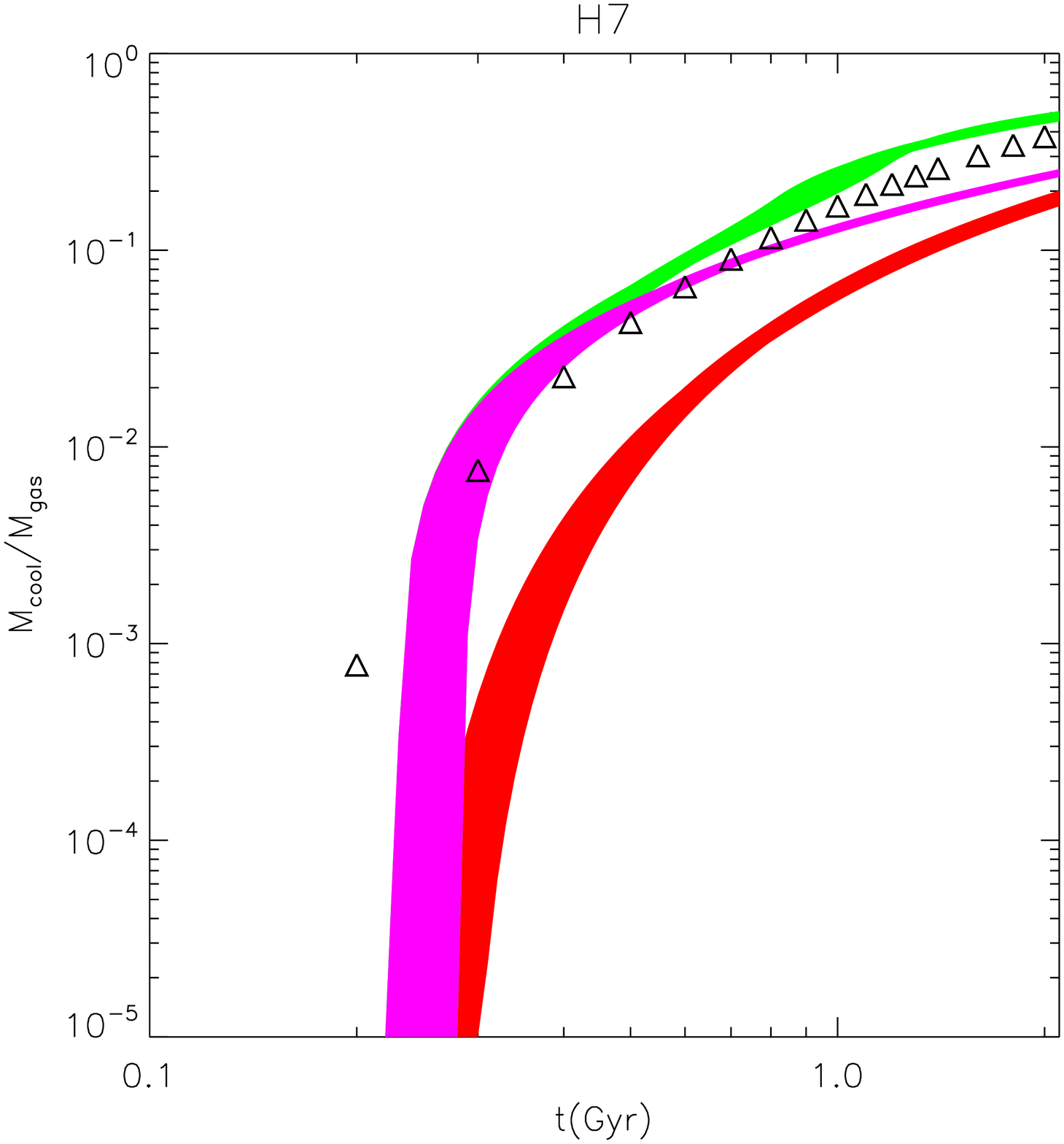,width=6cm}
\psfig{figure=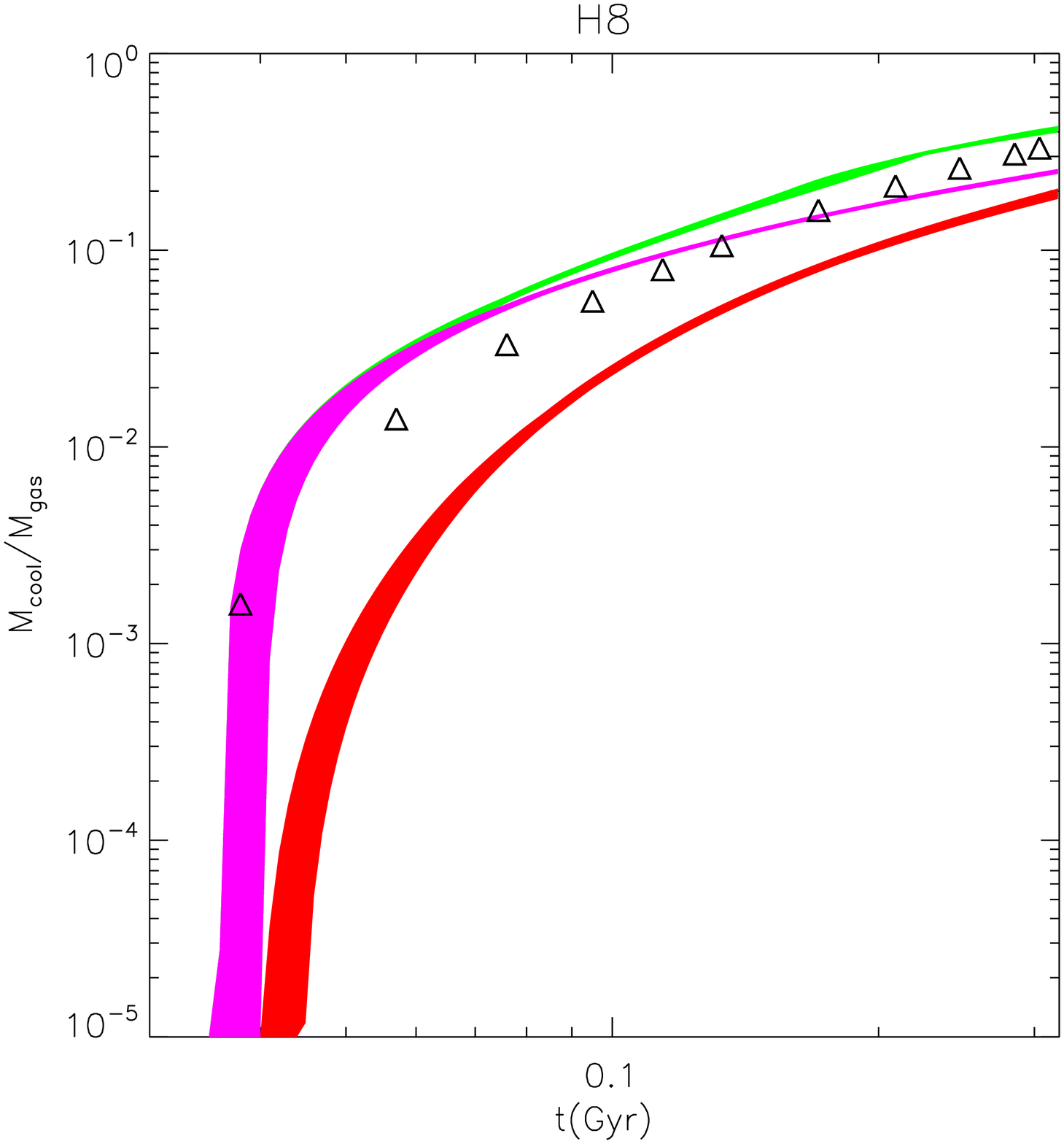,width=6cm}
\psfig{figure=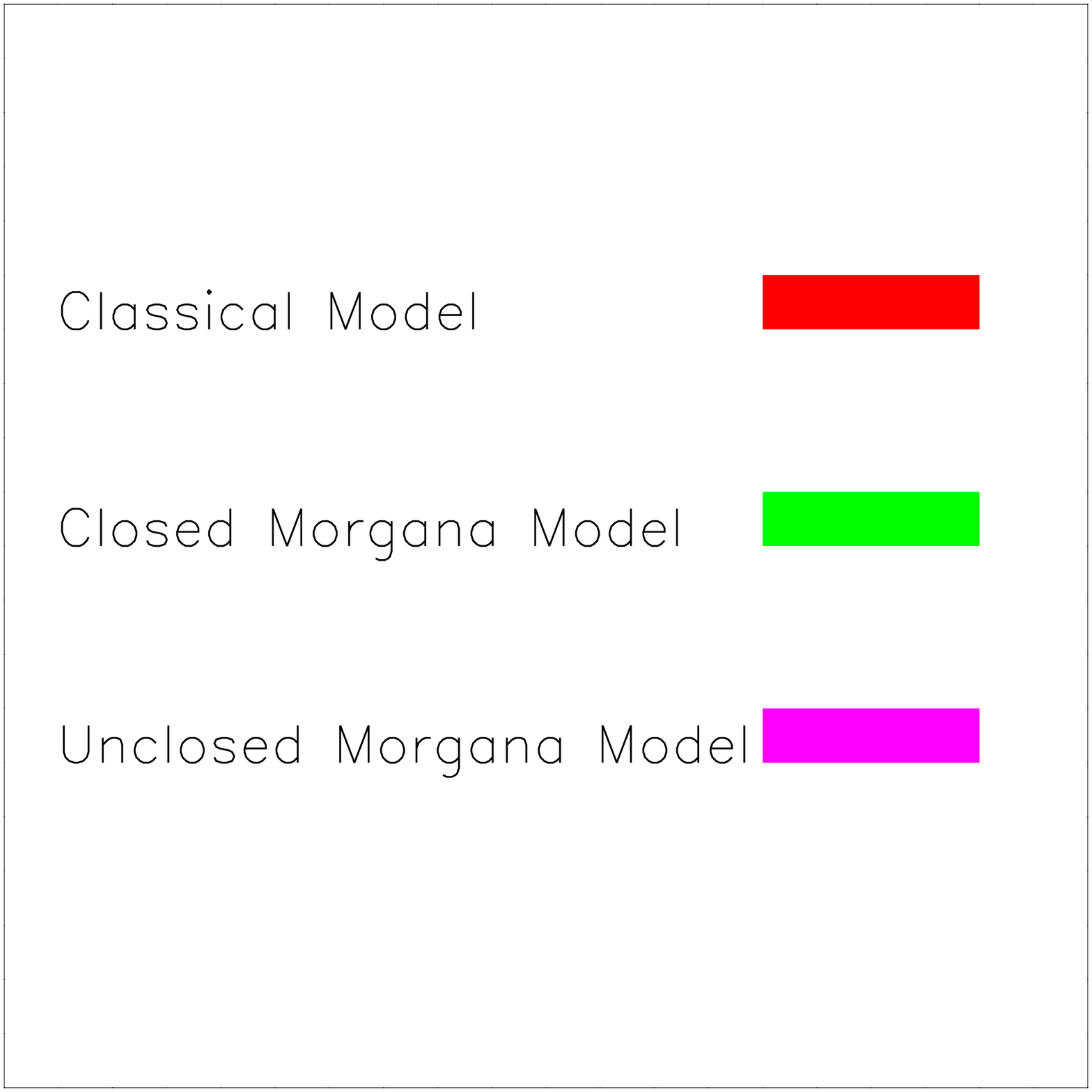,width=6cm}
}
}}
\end{center}
\caption{Evolution of the cooled gas fraction, $M_{\rm cool}$, for the
  eight simulated halos, in the radiative runs with star formation.
  Open triangles are the simulation results, while the shaded areas
  (in color in the online version) represent the prediction of the
  three different models. Classical model: darkest shade, red;
  unclosed {\sc morgana}: lightest shade, magenta; closed {\sc
  morgana}: middle shade, green). Each area represents the envelopes
  of hydrostatic models which provide a good fit to the initial
  conditions (see text).}
\label{fig:coolingmass}
\end{figure*}

As a main term of comparison between analytic models and simulations,
we use the evolution of the cooled mass fraction.  In Figure
\ref{fig:coolingmass} we plot the evolution of $M_{\rm cool}$ for all the
eight simulated halos from simulations and the predictions of the
different cooling models. As for the latter, each shaded area
represent the envelope of each model prediction for all the profiles
which provide a good fit to the initial conditions.

From these plots we infer the following conclusions.

\noindent
{\bf (i)} No gas is allowed to cool down to low temperatures
  before one central cooling time has elapsed, both
  in the classical and in the {\sc morgana} models. Afterwards,
  cooling takes place abruptly, giving rise to a sort of ``burst'' of
  star formation. This behaviour is consistent with simulation results, at
  least when the scatter in density and temperature, which are present
  in the initial conditions, is reduced. \\
{\bf (ii)} The classical cooling model (red area) always underpredicts
  the value of $M_{\rm cool}$. This underestimate is more severe at
  earlier times, and remains quite substantial, a factor of 2--3, even
  in the most evolved configurations. This effect is generally
  stronger for the halos having a lower concentration and/or higher
  mass, i.e. having longer cooling times (see Table 1).\\ 
{\bf (iii)} The unclosed {\sc morgana} model (magenta area) follows in
  a much closer way the cooled mass at early times and the fit is very
  good in most cases at epochs between one and a few central cooling
  times. In the most evolved configurations, the model underestimate
  the cooled mass, although the underestimate is sensibly reduced with
  respect to the classical model.\\
{\bf (iv)} The closed {\sc morgana} model behaves very similarly to the
  unclosed one for a few central cooling times. At later times, it
  predicts a larger value of $M_{\rm cool}$, providing a generally
  good fit to the simulation results for the most evolved
  configurations.

\section{Discussion}

The better performance of the unclosed {\sc morgana} model, with
respect to the classical model, in reproducing the evolution of the
cooled gas fraction from the simulations can be ascribed to the
following facts.

As explained in Section~2.4, the classical model relies on the strong
hypothesis that each gas shell cools in exactly one cooling time, with
this cooling time computed on the initial conditions. While this
hypothesis is valid when the first gas particles cool, it breaks down
soon later. This behaviour is indeed not unexpected.  The evolution of a
mass element in presence of cooling and adiabatic compression is given
by the following equation:
\begin{equation}
\dot{T}=T\left(-\frac{1}{t_{\rm cool}} +\frac{2}{3}\frac{\dot{\rho}}{\rho}\right)\,,
\label{eq:masselement}
\end{equation}
where $\rho(t)$ and $T(t)$ describe the evolution of density and
temperature of the gas element, and $t_{\rm cool}$ is the cooling time
computed on the actual density and temperature (not on the initial
conditions).  Under the assumptions that the temperature dependence of
the cooling time can be neglected, so that $t_{\rm cool}(t) = t_{\rm
c0} (\rho(t)/\rho_0)^{-1}$, and that pressure is constant during the
evolution, it is easy to solve equation (\ref{eq:masselement}) and
find that the time $t_{\rm tot}$ required for the mass element to cool
to $T=0$ coincides with $t_{\rm c0}$. Therefore, the basic assumption
of the classical model is satisfied in the case in which gas particles
cool at constant pressure.

On the other hand, Figure~\ref{fig:temptime} shows that gas
particles take most time to flow from their initial position to the
cooling region. Since cooling takes place in a pressure-supported way
during this time, their radiative losses are balanced by adiabatic
compression. As a result, the temperature of the gas particles has a
slow evolution while density increases significantly as they move
towards the cooling region.  This results in shallow and stable
temperature profiles, while density profiles become progressively
steeper (see Figure~\ref{fig:pressure}). The density increase turns
into enhanced radiative losses, thereby making the total cooling time,
$t_{\rm tot}$, shorter than the cooling time, $t_{\rm c0}$, computed
on the initial conditions.

The main assumption of the classical model is then clearly invalidated
by our simulations. Going back to the original proposal of this model
by \cite{White91}, the main justification was that the model is
roughly consistent with the exact self-similar solutions found by
\cite{Bert89}.  However, Bertschinger's self similar solutions give
cooling flows that are equal to:
\begin{equation}
\dot{M}_{\rm cool} = 4 \pi r^2\rho_g(r_{\rm C}) \frac{dr_{\rm C}}{dt} \times \mu_0\, ,
\label{eq:bert} \end{equation}
where the constant $\mu_0$ depends on the initial profile and can take
values ranging from $\sim0.1$ to $\sim2.5$ (see tables 1 and 2 in
Bertchinger's paper).  The classical model is recovered in the case
$\mu_0=1$.  In the Appendix we will consider the simple case of an
isothermal halo with power-law density profile $\rho_g\propto r^{-2}$.
In this case the unclosed {\sc morgana} model gives $t_{\rm tot}=0.5
t_{\rm c0}$ and a mass-deposition rate, $\dot{M}_{\rm cool}$, which is
equal to $\sqrt{2}$ times that of the classical model. On the other
hand, Bertschinger's solutions give cooling flows higher than the
classical value by a factor ranging from 1.304 to 1.190, depending on
the shape of the cooling function, thus implying total cooling times
$t_{\rm tot}$ shorter than $t_{\rm c0}$ by a factors ranging from 0.59
to 0.70.  Both {\sc morgana} and Bertschinger's self-similar solutions
predict that shallower power-law profiles give stronger cooling flows
and shorter total cooling times.  Therefore, while the unclosed {\sc
morgana} model agrees with the exact self-similar solution to within a
few tens per cent, the assumption that $t_{\rm tot}=t_{\rm c0}$ is
generally not correct.

More in detail, the difference between classical and unclosed {\sc
morgana} models at the onset of cooling is the consequence of the
flatness of the density profile in the inner regions of the gas.  This
can be shown by finding exact solutions of the classical and {\sc
morgana} models in the case of power-law density profiles,
$\rho_g(r)\propto r^{-\alpha}$, for an isothermal gas. These
calculations are reported in the Appendix and the results can be
summarized as follows.
{\bf (i)} Both the unclosed {\sc morgana} model and the classical one
  predict a self-similar cooling flow for $\alpha>3/2$, thus
  implying that the corresponding mass deposition rates are
  proportional to each other.
{\bf (ii)} The classical and unclosed {\sc morgana} models agree if
  $\alpha=3$; shallower profiles lead the unclosed {\sc morgana} model
  to predict shorter total cooling times and higher cooling rates.
{\bf (iii)} For profiles flatter than $\alpha=3/2$ the mass cooling
  flow of the unclosed {\sc morgana} model is dominated by the
  external regions.  The solution is not proportional to the classical
  one, the cooling flow is not self-similar and is roughly constant.
  Clearly, such a
  shallow profile will hold in the inner region of a realistic halo.

As a conclusion, the strict validity of the classical model is limited
to specific profiles and to self-similar flows. On the other hand, the
unclosed {\sc morgana} model, which relaxes the assumption on the
total cooling time of a gas shell, better reproduces the stronger
flows found in our simulations of isolated halos, which takes place
when the Lagrangian cooling radius sweeps the flat part of the density
profile.

The closed {\sc morgana} model further improves the agreement with the
simulations by increasing the fraction of cooled mass.  The main
reason for this increase is that, being always $r_{\rm M,ch}<r_{\rm
M}$, the smaller value of the cooling radius leads to an increase of
the density of the cooling shell, simply because the hot gas is
allowed to stay at smaller radii. This implies still shorter cooling
times and enhanced cooling flows. Another prediction of the closed
{\sc morgana} model is that the cooling radius $r_{\rm M,ch}$ is
stable after a quick transient. This prediction is in qualitative
agreement with the results of the simulations
(see Figure~\ref{fig:eul_temp}; the dotted lines give the position of
$r_{\rm M,ch}$). However, the size of the cooling region in the
simulation is affected by resolution, so this comparison cannot be
pushed to a quantitative level. The validity of this
model breaks as soon as the energy of the uncooled gas 
drops below the
virial value. In this case we still obtain a good match of the cooled
mass fraction by simply dropping the sound speed term in equation
(\ref{eq:drcool2}), which is responsible for
the shrinking of the cooling hole. 
For this reason, we consider the
closed {\sc morgana} model as an effective model, in that it takes
into account the shrinking of the cooling region caused by the
pressure from the hot gas just outside this region, without however
providing a formally rigorous description for this effect.

Using the predicted rough constancy of the cooling flow when the
central, shallow part of the gas profile is cooling, in Appendix B we
show that it is possible to give a very simple and remarkably accurate
prediction of the cooled mass as the result of a constant flow which
is given by a simple analytic formula, valid up to $\sim10$ central
cooling times.

\section{Conclusions}
We have presented a detailed analysis of cooling of hot gas in DM
halos, comparing the predictions of semi-analytic models with the
results of controlled numerical experiments of isolated NFW
halos with hot gas in hydrostatic equilibrium. Simulations have been
performed spanning a range of masses (from galaxy- to cluster-sized
halos), concentrations and redshift (from 0 to 2). Smaller halos at
higher redshift have not been simulated because the validity of the
assumption of a hydrostatic atmosphere is doubtful when the cooling
time is much shorter than the dynamical time.

We have considered the ``classical'' cooling model of \cite{White91},
used in most SAMs, and the model
recently proposed by \cite{Monaco07} within the {\sc morgana} code for
the evolution of galaxies and AGNs.  The main features of these models
can be summarized as follows.  The density and pressure profiles of
the gas are computed by solving the equation of hydrostatic
equilibrium in an NFW halo \citep{Suto98}.  The classical cooling
model assumes that each mass shell cools to low temperature exactly
after one cooling time $t_{\rm cool}(r)$, computed on the initial
conditions.  The cooling radius $r_{\rm C}$ is then the inverse of the
$t_{\rm cool}(r)$ function, and the cooled fraction is the fraction of
gas mass within $r_{\rm C}$.  The ``unclosed {\sc morgana}'' cooling
model computes the cooling rate of each mass shell, then integrates
over the contribution of all mass shells and follows the evolution of
the cooling radius assuming that the transition from hot to cold
phases is quick enough so that a sharp border in the density profile
of hot gas is always present.  This determines the evolution of the
cooling radius $r_{\rm M}$.  Moreover, to mimic the closure of the
``cooling hole'' due to the lack of pressure support at $r_{\rm M}$,
the cooling radius (now called $r_{\rm M,ch}$) is induced to close at
the sound speed. This defines the ``closed {\sc morgana}'' model.

Our main results can be summarized as follows.

\noindent
{\bf (i)} The classical cooling model systematically underestimates
the fractions of cooled mass. After about two central cooling times,
they are predicted to be about one order of magnitude smaller than
those found in simulations. Although this difference decreases with
time, after 8 central cooling times, when simulations are stopped, the
difference still amounts to a factor 2--3. This disagreement is
ascribed to the lack of validity of the assumption that each mass
shell takes one cooling time, computed on the initial conditions, to
cool to low temperature. Seen from the point of view of a mass
element, the time required by it to cool to low temperature is shorter
than the initial cooling time when density increases and temperature is
constant during cooling.  This is what happens to gas particles in the
simulations: they take most of time to travel from their initial
position toward the cooling region, at roughly constant temperature
and increasing density.  The disagreement is stronger when the cooling
gas comes from the shallow central region, in which case the cooling
flow is markedly not self-similar.

\noindent
{\bf (ii)} The unclosed {\sc morgana} model gives a much better fit of
the cooled mass fraction.  This is mostly due to the relaxation of the
assumption on the cooling time, mentioned in point (i).  This model
correctly predicts cooling flows which are stronger than the classical
model, by a larger amount for flatter gas density profiles.  In the
Appendix we show that the solution is not self-similar if the slope of
the density profile is shallower than $r^{-3/2}$. In this case cooling
is not dominated by the the shells just beyond the cooling radius but
the whole region for which the density profile is shallow contributes.

\noindent
{\bf (iii)} The closed {\sc morgana} model further improves the fit to
the simulation results on the evolution of the cooled mass fraction,
giving accurate results to within 20--50 per cent in all the
considered cases, after about 8 central cooling times.  This agreement is a
good reward for the increase of physical motivation of this model,
obtained at the modest price of letting the cooling radius close at
the local sound speed.  However, the closure of the cooling radius
must be halted at later times for the model to give realistic
results. In general, we consider the closed {\sc morgana} as a
successful effective model of cooling, rather than as a rigorous
physical model.

\noindent
{\bf (iv)} The cooling flow is well approximated by a constant flow,
for which we give a fitting formula in Appendix B, and which is valid
up to $\sim10$ central cooling times.

In the context of models of galaxy formation, cooling of hot
virialized gas is the starting point for all the astrophysical
processes involved in the formation of stars (and supermassive black
holes) and their feedback on the interstellar and intra-cluster media.
We find that the classical model, used in most SAMs, leads to a
significant underestimate of the cooled mass at early times. This
result is apparently at variance with previous claims, 
discussed in the Introduction, of an agreement
of models and simulations in predicting the cooled mass, 
(\citealt{Benson01,Helly03,Cattaneo07,Yoshida02}).
Given the much higher complexity of the cosmological initial
conditions used in such analyses, it is rather difficult to perform a
direct comparison with the results of our simulations of isolated
halos. Here we only want to stress the advantage of performing simple
and controlled numerical tests in order to study in detail how the
process of gas cooling takes place.

It is well possible that the different behaviour of simulations and
the classical cooling model is less apparent when the more complex
cosmological evolution is considered.  However, there is no doubt that
the results of our analysis are quite relevant for the comparison
between SAM predictions and observations. For instance,
\citep{Fontanot07} have recently shown that the {\sc morgana} model of
galaxy formation is able to reproduce the observed number counts of
sources in the sub-mm band by using the standard Initial Mass Function
(IMF) by \cite{Salpeter55}, without any need to resort to a
top-heavier IMF \citep{Baugh06}.  As argued by these authors, the bulk
of starbursts are driven by massive cooling flows, so this difference
is mostly due to the different cooling models.

\section*{Acknowledgments}
We wish to thank Volker Springel for having provided us with the
non-public version of GADGET-2 and Fabio Fontanot, Ian McCarthy,
Richard Bower and Klaus Dolag for many enlightening
discussions.  The simulations have been realized using the
super-computing facilities at the ``Centro Interuniversitario del
Nord-Est per il Calcolo Elettronico'' (CINECA, Bologna), with CPU time
assigned thanks to an INAF--CINECA grant and to an agreement between
CINECA and the University of Trieste. This work has been partially
supported by the INFN PD-51 grant and by a ASI/INAF grant for the
support of theory activity.

\bibliographystyle{mn2e}
\bibliography{biblio}

\begin{thebibliography}{}

\bibitem[\protect\citeauthoryear{{Baugh}}{{Baugh}}{2006}]{Baugh06}
{Baugh} C.~M.,  2006, Reports of Progress in Physics, 69, 3101

\bibitem[\protect\citeauthoryear{{Benson}, {Pearce}, {Frenk}, {Baugh} \&
  {Jenkins}}{{Benson} et~al.}{2001}]{Benson01}
{Benson} A.~J.,  {Pearce} F.~R.,  {Frenk} C.~S.,  {Baugh} C.~M.,    {Jenkins}
  A.,  2001, \mnras, 320, 261

\bibitem[\protect\citeauthoryear{{Bertschinger}}{{Bertschinger}}{1989}]{Bert89}
{Bertschinger} E.,  1989, \apj, 340, 666

\bibitem[\protect\citeauthoryear{{Cattaneo}, {Blaizot}, {Weinberg}, {Kere{\v
  s}}, {Colombi}, {Dav{\'e}}, {Devriendt}, {Guiderdoni} \& {Katz}}{{Cattaneo}
  et~al.}{2007}]{Cattaneo07}
{Cattaneo} A.,  {Blaizot} J.,  {Weinberg} D.~H.,  {Kere{\v s}} D.,  {Colombi}
  S.,  {Dav{\'e}} R.,  {Devriendt} J.,  {Guiderdoni} B.,    {Katz} N.,  2007,
  \mnras, 377, 63

\bibitem[\protect\citeauthoryear{{Cole}, {Lacey}, {Baugh} \& {Frenk}}{{Cole}
  et~al.}{2000}]{Cole00}
{Cole} S.,  {Lacey} C.~G.,  {Baugh} C.~M.,    {Frenk} C.~S.,  2000, \mnras,
  319, 168

\bibitem[\protect\citeauthoryear{{Fontanot}, {Monaco}, {Silva} \&
  {Grazian}}{{Fontanot} et~al.}{2007}]{Fontanot07}
{Fontanot} F.,  {Monaco} P.,  {Silva} L.,    {Grazian} A.,  2007, \mnras,
  accepted, 0, 0

\bibitem[\protect\citeauthoryear{{Gnedin}, {Kravtsov}, {Klypin} \&
  {Nagai}}{{Gnedin} et~al.}{2004}]{Kravtsov04}
{Gnedin} O.~Y.,  {Kravtsov} A.~V.,  {Klypin} A.~A.,    {Nagai} D.,  2004, \apj,
  616, 16

\bibitem[\protect\citeauthoryear{{Helly}, {Cole}, {Frenk}, {Baugh}, {Benson},
  {Lacey} \& {Pearce}}{{Helly} et~al.}{2003}]{Helly03}
{Helly} J.~C.,  {Cole} S.,  {Frenk} C.~S.,  {Baugh} C.~M.,  {Benson} A.,
  {Lacey} C.,    {Pearce} F.~R.,  2003, \mnras, 338, 913

\bibitem[\protect\citeauthoryear{{Hernquist}}{{Hernquist}}{1993}]{Hernquist93}
{Hernquist} L.,  1993, \apjs, 86, 389

\bibitem[\protect\citeauthoryear{{Kauffmann}, {White} \&
  {Guiderdoni}}{{Kauffmann} et~al.}{1993}]{Kauffmann93}
{Kauffmann} G.,  {White} S.~D.~M.,    {Guiderdoni} B.,  1993, \mnras, 264, 201

\bibitem[\protect\citeauthoryear{{Komatsu} \& {Seljak}}{{Komatsu} \&
  {Seljak}}{2001}]{Komatsu01}
{Komatsu} E.,  {Seljak} U.,  2001, \mnras, 327, 1353

\bibitem[\protect\citeauthoryear{{Menci}, {Fontana}, {Giallongo} \&
  {Salimbeni}}{{Menci} et~al.}{2005}]{Menci05}
{Menci} N.,  {Fontana} A.,  {Giallongo} E.,    {Salimbeni} S.,  2005, \apj,
  632, 49

\bibitem[\protect\citeauthoryear{{Monaco}, {Fontanot} \& {Taffoni}}{{Monaco}
  et~al.}{2007}]{Monaco07}
{Monaco} P.,  {Fontanot} F.,    {Taffoni} G.,  2007, \mnras, 375, 1189

\bibitem[\protect\citeauthoryear{{Nagai} \& {Kravtsov}}{{Nagai} \&
  {Kravtsov}}{2005}]{Nagai05}
{Nagai} D.,  {Kravtsov} A.~V.,  2005, \apj, 618, 557

\bibitem[\protect\citeauthoryear{{Nagamine}, {Cen}, {Hernquist}, {Ostriker} \&
  {Springel}}{{Nagamine} et~al.}{2004}]{Nagamine04}
{Nagamine} K.,  {Cen} R.,  {Hernquist} L.,  {Ostriker} J.~P.,    {Springel} V.,
   2004, \apj, 610, 45

\bibitem[\protect\citeauthoryear{{Navarro}, {Frenk} \& {White}}{{Navarro}
  et~al.}{1997}]{NFW}
{Navarro} J.~F.,  {Frenk} C.~S.,    {White} S.~D.~M.,  1997, \apj, 490, 493

\bibitem[\protect\citeauthoryear{{Power}, {Navarro}, {Jenkins}, {Frenk},
  {White}, {Springel}, {Stadel} \& {Quinn}}{{Power} et~al.}{2003}]{Power03}
{Power} C.,  {Navarro} J.~F.,  {Jenkins} A.,  {Frenk} C.~S.,  {White} S.~D.~M.,
   {Springel} V.,  {Stadel} J.,    {Quinn} T.,  2003, \mnras, 338, 14

\bibitem[\protect\citeauthoryear{{Romeo}, {Portinari} \&
  {Sommer-Larsen}}{{Romeo} et~al.}{2005}]{Romeo05}
{Romeo} A.~D.,  {Portinari} L.,    {Sommer-Larsen} J.,  2005, \mnras, 361, 983

\bibitem[\protect\citeauthoryear{{Salpeter}}{{Salpeter}}{1955}]{Salpeter55}
{Salpeter} E.~E.,  1955, \apj, 121, 161

\bibitem[\protect\citeauthoryear{{Saro}, {Borgani}, {Tornatore}, {Dolag},
  {Murante}, {Biviano}, {Calura} \& {Charlot}}{{Saro} et~al.}{2006}]{Saro06}
{Saro} A.,  {Borgani} S.,  {Tornatore} L.,  {Dolag} K.,  {Murante} G.,
  {Biviano} A.,  {Calura} F.,    {Charlot} S.,  2006, \mnras, 373, 397

\bibitem[\protect\citeauthoryear{{Somerville} \& {Primack}}{{Somerville} \&
  {Primack}}{1999}]{Somerville99}
{Somerville} R.~S.,  {Primack} J.~R.,  1999, \mnras, 310, 1087

\bibitem[\protect\citeauthoryear{{Springel}}{{Springel}}{2005}]{Springel05gad}
{Springel} V.,  2005, \mnras, 364, 1105

\bibitem[\protect\citeauthoryear{{Springel}, {Frenk} \& {White}}{{Springel}
  et~al.}{2006}]{Springel06}
{Springel} V.,  {Frenk} C.~S.,    {White} S.~D.~M.,  2006, \nat, 440, 1137

\bibitem[\protect\citeauthoryear{{Springel} \& {Hernquist}}{{Springel} \&
  {Hernquist}}{2002}]{SprHern02}
{Springel} V.,  {Hernquist} L.,  2002, \mnras, 333, 649

\bibitem[\protect\citeauthoryear{{Sutherland} \& {Dopita}}{{Sutherland} \&
  {Dopita}}{1993}]{SD93}
{Sutherland} R.~S.,  {Dopita} M.~A.,  1993, \apjs, 88, 253

\bibitem[\protect\citeauthoryear{{Suto}, {Sasaki} \& {Makino}}{{Suto}
  et~al.}{1998}]{Suto98}
{Suto} Y.,  {Sasaki} S.,    {Makino} N.,  1998, \apj, 509, 544

\bibitem[\protect\citeauthoryear{{Tornatore}, {Borgani}, {Springel},
  {Matteucci}, {Menci} \& {Murante}}{{Tornatore} et~al.}{2003}]{Tornatore03}
{Tornatore} L.,  {Borgani} S.,  {Springel} V.,  {Matteucci} F.,  {Menci} N.,
  {Murante} G.,  2003, \mnras, 342, 1025

\bibitem[\protect\citeauthoryear{{White} \& {Frenk}}{{White} \&
  {Frenk}}{1991}]{White91}
{White} S.~D.~M.,  {Frenk} C.~S.,  1991, \apj, 379, 52

\bibitem[\protect\citeauthoryear{{Yoshida}, {Stoehr}, {Springel} \&
  {White}}{{Yoshida} et~al.}{2002}]{Yoshida02}
{Yoshida} N.,  {Stoehr} F.,  {Springel} V.,    {White} S.~D.~M.,  2002, \mnras,
  335, 762

\end{thebibliography}

\appendix
\section{Model solutions for power-law profiles}
In this Appendix we will discuss the cases in which the unclosed {\sc
morgana} cooling model provides self-similar solutions and how these
solutions compare to the exact self-similar solutions by
\cite{Bert89} and to the classical cooling model by \cite{White91}. To
this purpose,
we analytically solve equation (\ref{eq:morganacooling}) for the
evolution of the cooled gas mass $M_{\rm cool}$ and equation
(\ref{eq:drcool1}) for the evolution of the cooling radius $r_{\rm M}$ in
the case of an isothermal power-law density profile:
\begin{equation}
\rho_g(r) = \rho_{gV} \left( \frac{r}{r_{200}} \right)^{-\alpha}\,.
\end{equation}
Here $\rho_{gV}$ is the density at the virial radius $r_{200}$, while
temperature is fixed to the virial one.  In the following we will make
the approximation that we can neglect the dependence of the cooling
time on temperature:
\begin{equation}
t_{\rm cool}(r) = t_{cV} \left( \frac{r}{r_{200}} \right)^{\alpha} \, ,
\end{equation}
where $t_{cV}$ is the initial cooling time of the gas at the virial
radius.  
Let $r_{\rm M}$ be the cooling radius of the unclosed Morgana model.
The mass cooling flow (equation~\ref{eq:morganacooling}) is:
\begin{equation}
\dot{M}_{\rm cool} = 
\frac{4\pi r_{200}^3 \rho_{gV}}{t_{cV}} \int_{r_{\rm M}/r_{200}}^1 \!\!\!\!\!\!x^{-2\alpha+2}dx \,,
\end{equation}
where $x\equiv r/r_{200}$. If $\alpha\ne 3/2$ the solution is:
\begin{equation}
\dot{M}_{\rm cool} = \frac{4\pi r_{200}^3 \rho_{gV}}{t_{cV}} \frac{1}{2\alpha-3} \left[
\left( \frac{r_{\rm M}}{r_{200}}\right)^{-(2\alpha-3)} -1 \right] \, .
\label{eq:ap_dMcool}
\end{equation}
If $\alpha>3/2$ and $r_{\rm M} \ll r_{200}$ then the equation for
the Morgana cooling radius $r_{\rm M}$ becomes (equation~\ref{eq:drcool1}):
\begin{equation}
\dot{r}_M = \frac{\dot{M}_{\rm cool}}{4\pi r_{\rm M}^2\rho_g(r_{\rm M})} = 
\frac{r_{200}}{(2\alpha-3)t_{cV}}
\left(\frac{r_{\rm M}}{r_{200}}\right)^{1-\alpha}\, ,
\label{eq:ap_drcool} \end{equation}
whose solution is
\begin{equation}
r_{\rm M} = r_{200} \left( \frac{\alpha}{2\alpha-3} \frac{t}{t_{cV}} \right)^{1/\alpha}\,.
\end{equation}
Substituting this solution into equation (\ref{eq:ap_dMcool}) and neglecting
the $-1$ term in parentheses we obtain
\begin{equation}
\dot{M}_{\rm cool} = \frac{4\pi r_{200}^3 \rho_{gV}}{t_{cV}} \frac{1}{2\alpha-3}
\left(\frac{\alpha}{2\alpha-3} \frac{t}{t_{cV}} \right)^{-(2\alpha-3)/\alpha}\, .
\end{equation}

An analogous computation can be performed for the classical cooling
model. From the definition of equation (\ref{eq:coolingradius})
for the classical cooling radius, we obtain:
\begin{equation}
r_{\rm C}(t) = r_{200} \left(\frac{t}{t_{cV}}\right)^{1/\alpha}\,,
\end{equation}
which gives
\begin{equation}
\dot{M}_{\rm classic} =\frac{4\pi r_{200}^3 \rho_{gV}}{\alpha t_{cV}} \left(\frac{t}{t_{cV}}\right)^{(3-2\alpha)/\alpha} 
\end{equation}
for the mass deposition rate. 
Therefore, the comparison of the two models leads to
\begin{equation}
\dot{M}_{\rm cool} = \left(\frac{\alpha}{2\alpha-3}\right)^{(3-\alpha)/\alpha} 
\times \dot{M}_{\rm classic} \, .
\end{equation}
In the case of $\alpha=2$ we obtain $\dot{M}_{\rm cool} = \sqrt{2}
\dot{M}_{\rm classic}$. Since the mass deposition rate from the {\sc
morgana} model is proportional to the classical one, cooling is
self-similar and can be compared to the solution of \cite{Bert89}.
The latter depends on the (power-law) temperature dependence of the
cooling function, $\Lambda\propto T^\lambda$. To have a cooling time
independent of temperature, as assumed in the {\sc morgana} model, we
should compare to the solution for $\lambda=1$, which is not given by
Bertschinger because it is of no physical relevance. Going from
$\lambda=-1$ to $\lambda=0.5$ (the values considered by Bertschinger)
the ratio between the values of $\dot{M}_{\rm cool}$ predicted by the
self-similar solution and by the classical model grows from 1.190 to
1.304.  This compares well to our prediction of $\sqrt{2}$.  We then
conclude that the unclosed {\sc morgana} model is roughly consistent
with Bertschinger's exact self-similar solution within a few tens per
cent in the case of an isothermal gas density profile with $\alpha=2$.

The total cooling time $t_{\rm tot}$, i.e. the time required by a
shell to cool to very small temperature, is defined as the inverse of
the $r_{\rm M}(t)$ function.  The unclosed {\sc morgana} model gives:
\begin{equation}
t_{\rm tot}=\frac{2\alpha-3}{\alpha} t_{cV} \left(\frac{r}{r_{200}}\right)^{\alpha}
= \frac{2\alpha-3}{\alpha} t_{\rm cool}(r)\,,
\end{equation}
while the classical model assumes $t_{\rm tot}= t_{\rm cool}(r)$.
Then, complete cooling is predicted by the {\sc morgana} model to take
less than one cooling time (half of it if $\alpha=2$) whenever
$\alpha<3$. Clearly, the classical and unclosed {\sc morgana} model
give the same cooling flow and total cooling time for an isothermal
profile with $\alpha=3$.

If $\alpha<3/2$, the $-1$ term in equation (\ref{eq:ap_dMcool})
becomes dominant. As a consequence, the cooling flow is not
proportional to the classical one, so the solution is not
self-similar, and cannot be compared to Bertschinger's solutions.  In
fact, in a shallow profiles cooling is dominated by the external
regions and the flow is roughly constant as a first
approximation.  However, in realistic cases a shallow profiles will
be valid only within some reference scale radius $r_{\rm ref}$, and
cooling will proceed very quickly until the cooling radius $r_{\rm M}$
has reached $r_s$.

\section{A simple analytic expression for the cooling flow}

\begin{figure*}
\centerline{
    \psfig{figure=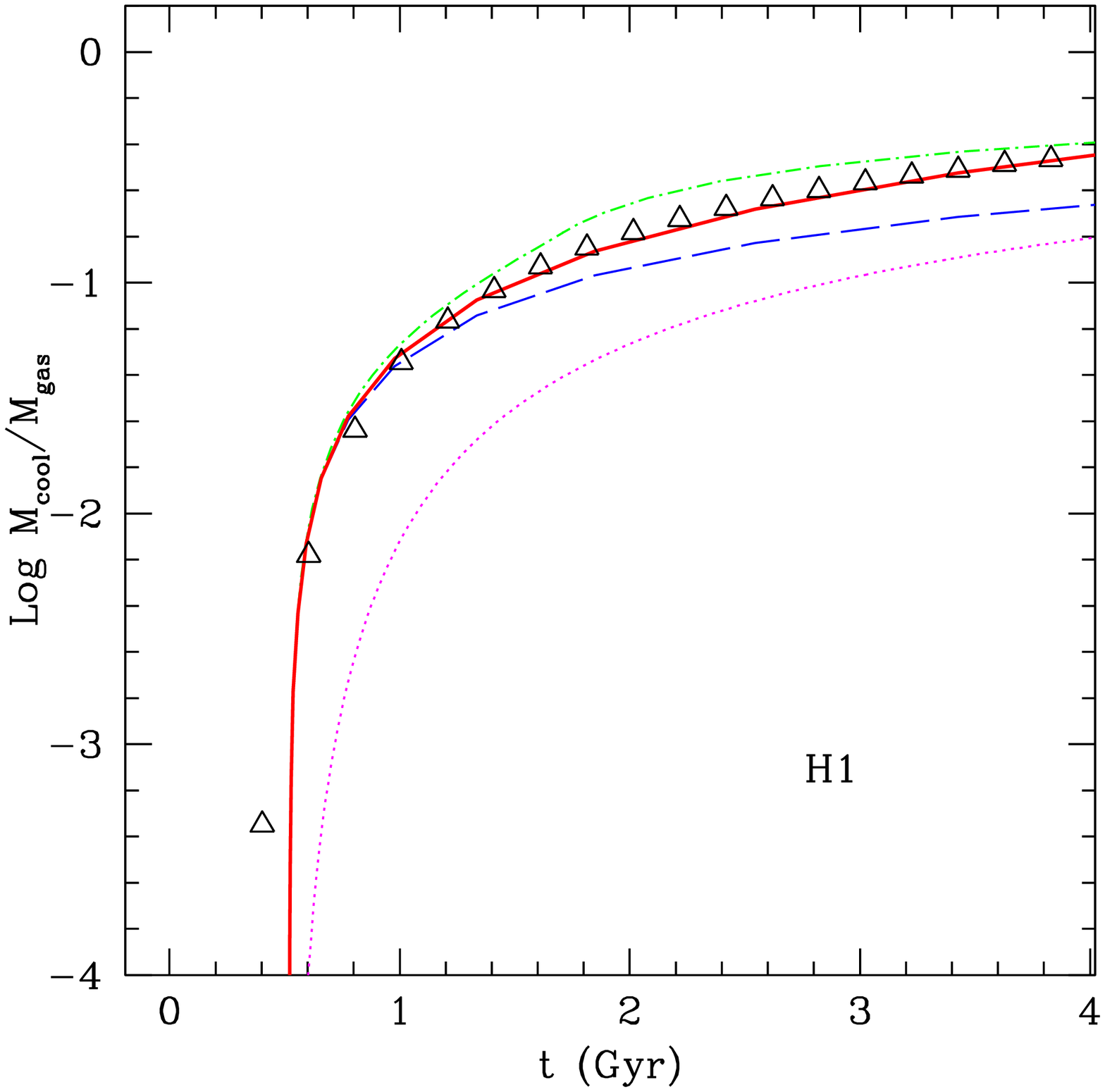,width=8cm}
    \psfig{figure=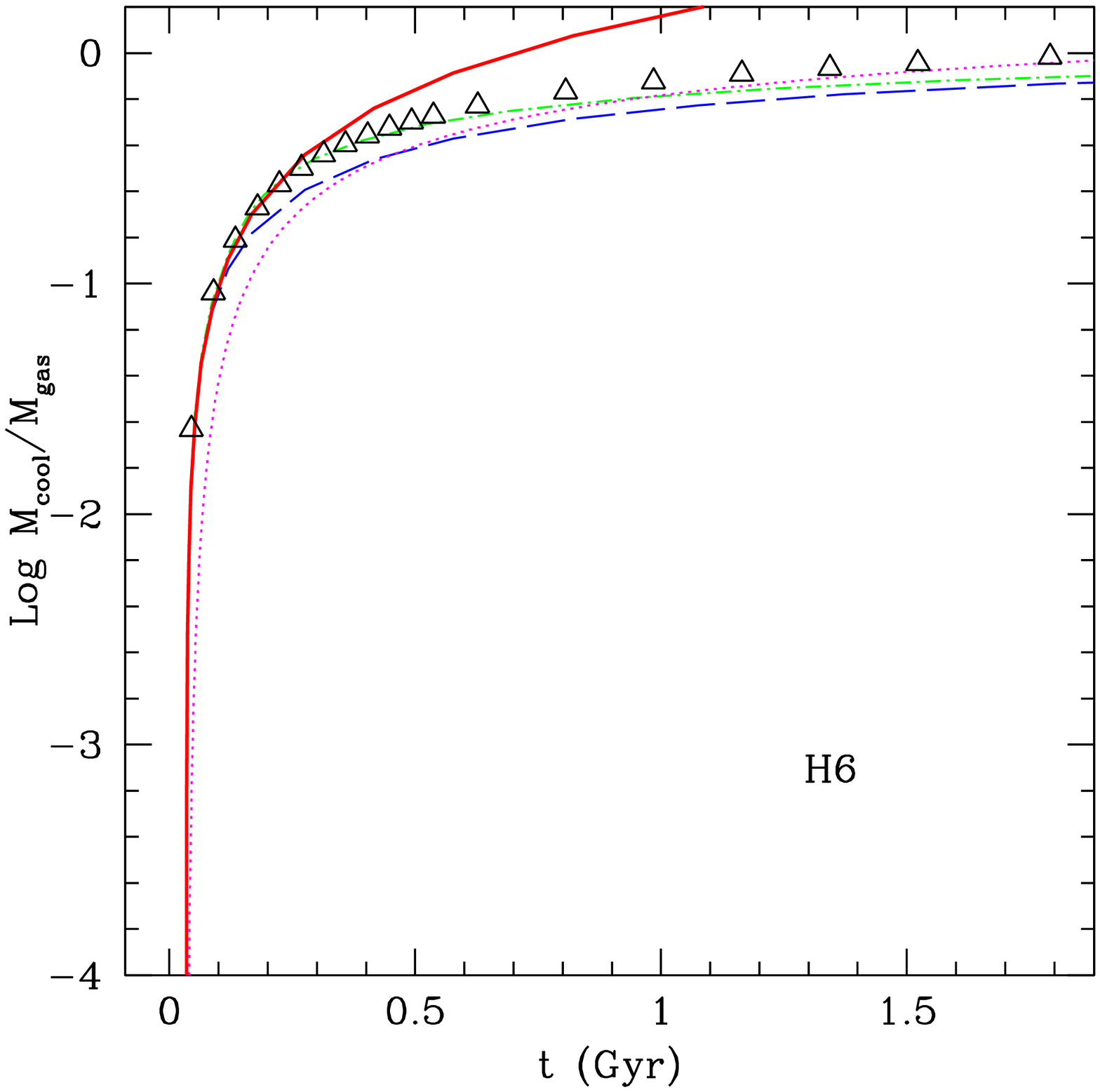,width=8cm}}
  \caption{Simple approximation to the cooled mass fraction obtained
  by integrating in time the constant cooling flow of
  equation~\ref{eq:best_guess}, shown for the H1 and H6 simulations
  (the others show a similar degree of agreement).  Dotted (magenta),
  dashed (blue) and dot-dashed (green) lines give the predictions of
  the classical, unclosed and closed {\sc morgana} models.  The thick
  continuous line gives the simple analytic fit. For sake of
  completeness, the models are obtained using $\gamma_p=1.20$ and
  thermal energy equal to 1.18 times the virial energy (H1), and
  $\gamma_p=1.21$ and energy equal to 1.15 the virial energy
  (H6). These values are at the centre of the intervals used for the
  models shown in figure~\ref{fig:coolingmass}.}
\label{fig:guess}
\end{figure*}

The last paragraph of Appendix A shows that, as long as the
cooling radius sweeps the region where the gas density profile is
shallower than $r^{-3/2}$, the cooling flow is roughly constant.
Indeed, we find that both simulations and the {\sc morgana} results
can be fit with a constant cooling flow.  To find an analytic
approximation for it, we start from computing the reference radius
$r_{\rm ref}$ at which $d\ln\rho_g/d\ln r=-3/2$.  The function
$\ln\rho_g$ (equation~\ref{eq:hydro_sol}) is Taylor-expanded around
$r_s$:

\begin{eqnarray}
\rho_g(r_s)&=&\rho_{g0}\left[1-a\left(1-\ln 2\right)\right]^{1/(\gamma_p-1)} \\
\frac{d\ln\rho_g}{d\ln r}(r_s) &=& \frac{a}{\gamma_p-1}\,\frac{1/2-\ln 2}{1-a(1-\ln 2)} \\
\frac{d^2\ln\rho_g}{d\ln r^2}(r_s) &=& \frac{a}{\gamma_p-1}\,\frac{\ln 2-3/4+a(2-3\ln 2)/4}{[1-a(1-\ln 2)]^2}\,.
\end{eqnarray}

\begin{eqnarray}
\lefteqn{\ln\rho_g(r) \simeq \ln \rho_g(r_s) }\\
&& + \ln\left(\frac{r}{r_s}\right) \frac{d\ln\rho_g}{d\ln r}(r_s) + \frac{1}{2} 
\left[\ln\left(\frac{r}{r_s}\right)\right]^2 
\frac{d^2\ln\rho_g}{d\ln r^2}(r_s) + \ldots \nonumber
\end{eqnarray}

\noindent
The reference radius is then:

\begin{equation}
r_{\rm ref} \simeq r_s \exp\left(\frac{-3/2-d\ln\rho_g/d\ln r(r_s)}{d^2\ln\rho_g/d\ln r^2(r_s)}\right)
\label{eq:reference}\end{equation}

A guess for the reference cooling flow can then be obtained as
$\dot{M} \simeq 4\pi r_{\rm ref}^3 \rho_g(r_{\rm ref})/t_{\rm
cool}(r_{\rm ref})$.  This guess has been compared to the value of the
cooling flow in the {\sc morgana} unclosed and closed models, averaged
over the time interval from 1 to 3 central cooling times and over the
two models.  We find systematic differences that are removed by
adopting a correction function $f(M_H,c_{\rm nfw})$ of halo mass and
concentration. Let's call:

\begin{eqnarray}
A &=& 1.95-\left(\frac{1.41\times10^{10}}{M_H}\right)^{0.3} \\
B &=& \max\left\{ 0.09\left[1-\left(\frac{M_H}{1.26\times10^{15}}\right)^{0.3}\right] , 0.06\right\} \\
C &=& -0.043\left(B-0.085\right)+0.0014
\end{eqnarray}

\noindent
then the correction function $f$ is found to be:

\begin{equation}
f(M_H,c_{\rm nfw})= A+(c_{\rm  nfw}-10)B
\end{equation}

\noindent
if $c_{\rm nfw}<=10$, and

\begin{equation}
f(M_H,c_{\rm nfw})= A+(c_{\rm  nfw}-10)B + (c_{\rm  nfw}-10)^2 C
\end{equation}

\noindent
if $c_{\rm nfw}>10$.  The approximated cooling flow is then:

\begin{equation}
\dot{M}_{\rm approx} \simeq f(M_H,c_{\rm nfw}) \times \frac{4\pi r_{\rm ref}^3
\rho_g(r_{\rm ref})}{t_{\rm cool}(r_{\rm ref})}
\label{eq:best_guess}\end{equation}

\noindent
The dependence on cosmology is hidden in the dependence on redshift
and concentration of $r_{\rm ref}$ and the function $f(M_H,c_{\rm nfw})$.

This simple analytical model gives a very good fit to all the
simulations shown in this paper, at least at early times.  For sake of
brevity, we show here (figure~\ref{fig:guess}) only the reference H1
case and the worst-case H6 simulation.  In the latter case the
fraction of cooled mass approaches unity, and the approximation of a
constant cooling flow breaks after $\sim10$ central cooling times.  It
is very simple to truncate the constant cooling flow by imposing that
the cooled mass does not overshoot the gas mass.  However, a precise
modeling of these late phases is immaterial, as feedback from star
formation and AGNs would clearly regulate the dynamics of cooling, so
any reasonable truncation would work equally well.

\end{document}